\pdfoutput=1
\documentclass[a4paper,12pt]{article}
\usepackage{amsfonts}
\usepackage{mathrsfs}
\usepackage{amsmath}
\usepackage{amssymb}
\usepackage[medium]{titlesec}
\usepackage{bm}
\usepackage{cite}
\usepackage[normalem]{ulem}
\usepackage{extarrows}

\usepackage{slashed}
\usepackage{isodateo}
\usepackage{graphicx}
\usepackage{xcolor}
\usepackage[bookmarksnumbered=true,bookmarksopen=true]{hyperref}
 \hypersetup{colorlinks,%
             linkcolor=[rgb]{0,0.3,0.6}, %
             citecolor=[rgb]{0,0.3,0.6}, %
             urlcolor=[rgb]{0,0.3,0.6}}
\usepackage[hmargin=.7in,vmargin=1.1in]{geometry}
\usepackage{indentfirst}
\newcommand{\FR}[2]{\displaystyle\frac{\,{#1}\,}{#2}}

\newcommand{\n}{\nonumber}

\graphicspath{{fig/}}

\def\bge{\begin{equation}}
\def\ede{\end{equation}}

\def\bga{\begin{aligned}}
\def\eda{\end{aligned}}
\def\bgp{\begin{pmatrix}}
\def\edp{\end{pmatrix}}
\def\bgs{\begin{subequations}}
\def\eds{\end{subequations}}

\def\di{{\mathrm{d}}}

\def\mb{\mathbf}

\def\la{\langle}\def\ra{\rangle}

\def\to{\rightarrow}

\def\ga{\gamma}
\def\de{\delta}
\def\ep{\epsilon}

\def\psidot{\dot\psi}

\begin{document}

\title{\Large\textbf{An Efficient Signal to Noise Approximation for Eccentric Inspiraling Binaries}}
\author{$^{a}$\,Lisa Randall\footnote{randall@g.harvard.edu},~~~~$^{b}$\,Alexandra Shelest\footnote{alexandra.shelest@epfl.ch},~~~~$^{c}$\,Zhong-Zhi Xianyu\footnote{zxianyu@tsinghua.edu.cn}\\[2mm]
\normalsize{$^a$\,\emph{Department of Physics, Harvard University, 17 Oxford St., Cambridge, MA 02138, USA}}\\[2mm]
\normalsize{$^b$\,\emph{Institute of Physics, School of Basic Sciences, EPFL, 1015 Lausanne, Switzerland}}\\[2mm]
\normalsize{$^c$\,\emph{Department of Physics, Tsinghua University, Beijing 100084, China}}}

\date{}
\maketitle

\begin{abstract}

Eccentricity has emerged as a potentially useful tool for helping to identify the origin of black hole mergers. However, owing to the large number of harmonics required to compute the amplitude of an eccentric signal, eccentric templates can be computationally very expensive, making statistical analyses to distinguish distributions from different formation channels very challenging. In this paper, we outline a method for estimating the signal-to-noise ratio for inspiraling binaries at lower frequencies such as those proposed for LISA and DECIGO. Our approximation can be useful more generally for any quasi-periodic sources. We argue that surprisingly, the signal-to-noise ratio evaluated at or near the peak frequency (of the power) is well approximated by using a constant noise curve, even if in reality the noise strain has power law dependence. We furthermore improve this initial estimate over our previous calculation to allow for frequency-dependence in the noise to expand the range of eccentricity and frequency over which our approximation applies. We show how to apply this method to get an answer accurate to within a factor of two over almost the entire projected observable frequency range. We emphasize this method is not a replacement for detailed signal processing. The utility lies chiefly in identifying theoretically useful discriminators among different populations and providing fairly accurate estimates for how well they should work. This approximation can furthermore be useful for narrowing down parameter ranges in a computationally economical way when events are observed. We furthermore show a distinctive way to identify events with extremely high eccentricity where the signal is enhanced relative to naïve expectations on the high frequency end.
\end{abstract}

\newpage

\section{Introduction} 

The LIGO/Virgo detections of coalescing black hole binaries (BBHs) marked the dawn of gravitational-wave (GW) astronomy \cite{LIGOScientific:2018jsj,LIGOScientific:2018mvr}. With increasing statistics from the ongoing run of LIGO/Virgo, we expect to learn many properties of stellar-mass BBHs. One of the important open questions is the formation channel of these black hole pairs \cite{Dominik:2012kk,Dominik:2013tma,Banerjee:2016ths,Samsing:2013kua,Samsing:2017xmd,Samsing:2018isx,Samsing:2019dtb,Antonini:2012ad,Hamers:2018hxv,Hoang:2017fvh,Fragione:2019dtr,Randall:2017jop,Randall:2018nud,Randall:2018lnh,Silsbee:2016djf,Randall:2019sab,Deme:2020ewx}.

Observing BBHs at  lower frequencies in the millihertz range with LISA (and sub-Hertz range with future detectors such as DECIGO)  could provide far more powerful information about the formation channels.  Measurements of the orbital eccentricity of BBHs \cite{Nishizawa:2016jji,Nishizawa:2016eza} in particular can help distinguish among populations.  Isolated BBHs typically possess little eccentricity while dynamically formed BBHs could have observably large eccentricity \cite{Banerjee:2016ths,Samsing:2013kua,Samsing:2017xmd,Samsing:2018isx,Samsing:2019dtb,Antonini:2012ad,Hamers:2018hxv,Hoang:2017fvh,Fragione:2019dtr,Randall:2017jop,Randall:2018nud,Randall:2018lnh,Silsbee:2016djf}.  Measuring eccentricity at different frequencies in  LIGO-Virgo-KAGRA (LVK), LISA, and DECIGO  will be an even more powerful method to identify formation channels.

However, the gravitational waves (GWs) radiated from inspiraling eccentric binaries possess multiple harmonics and the width of the spectrum increases with the eccentricity \cite{Barack:2003fp}. This complicates the calculation of the signal-to-noise ratio (SNR) as one has to sum over those many harmonics. The spread of frequencies makes the calculation of the average SNR more challenging in that one can no longer just calculate the signal and divide it by a fixed noise.
This is especially problematic for large eccentricity since one needs to evaluate many Bessel functions at large orders, which is computationally expensive. Therefore it would be good to have a formula relating the SNR of an eccentric binary to the SNR of a more tractable system.

In this paper we elaborate and extend 
 the simplified calculation  for the  SNR of eccentric binaries of \cite{Randall:2019znp}, where two of us showed how to relate the SNR of an eccentric binary to that of a circular binary with the same masses and peak frequency $f_p$, the frequency of the peak harmonic in the GW power spectrum.   We proposed an approximating formula that assumed a constant noise curve, using both an analytical argument at large $e$ and a numerical fit. Treating the SNR $\varrho(f_p,e)$ as a function of the peak frequency $f_p$ and the eccentricity $e$, we showed
\bge
  \varrho(f_p,e)/\varrho(f_p,e=0)\propto (1-e^2)^{3/4}
\ede
for large $e$. We then provided a full formula working for all $e\in [0,1)$  through a numerical fit.

For noise curves that decrease with frequency (or ones that decrease over the range of the signal), approximating the noise curve as a constant yields a partial cancellation between the frequencies that are lower and higher than the peak frequency. The corrections to our approximation would appear in the second derivative, which means it depends on the spread of frequencies divided by the peak frequency. So long as eccentricity is sufficiently far from unity, this will be a small(ish) number. Near the turnaround of the noise curve, there is less variation in noise from the get-go so the approximation there also works very well. Here we  show that for eccentricities less than 0.9, this approximation works rather well (at least at the level of 50\% accuracy and often better). This accuracy should be adequate for distinguishing among population distributions.

However, the partial cancellation does not necessarily work for the rising part of the noise curve because of the $1/n$ factor in the amplitude of the $n$'th harmonic, which manifests itself as a secondary peak in the spectrum at low frequency where the noise can be relatively low, as can be seen in Fig.\;\ref{spectrum}. 
 We deal with this by comparing the emission from an eccentric binary to the known (and more simple to calculate) emission of a circular binary emitting at the peak frequency determined from the eccentric system of interest but we allow for a power law dependence on frequency for the noise.   

Note that SNR involves the ratio of amplitude to noise, whereas the peak frequency is evaluated for the emitted power.  The difference is compensated by powers of $n$ (where $n$ labels the harmonic) and we show how to incorporate the $n$-dependence to get a simple formula. We show that our result is better than anticipated.  We derive a fairly precise formula for $\varrho(f_p,e)/\varrho(f_p,e=0)$ for all $e\in [0,1)$ without assuming any particular limit or using any numerical fit. We derive  formulae for different power laws of the noise curve. In particular, we consider three interesting cases: $S_N(f)\propto (f^0,f^{-4},f^2)$. These correspond to realistic noise curves such as LISA or DECIGO at the minimum noise, lower frequency, and higher frequency. The noise curves as used at this point have essentially this form (with some small deviations for LISA).  Of course real noise curves might have features that will have to be taken into account. But given the noise curves in the literature \cite{Klein:2015hvg,Yagi:2011wg}, our approximation works very well.

In this paper we clarify the original method, check in detail how well it should work in LISA and DECIGO, and extend the analysis to power law noise curves
 to show how accuracy can be improved even further, extending our analysis to higher frequencies as well.
 We show that for periodic signals, taking the sum over harmonics near peak frequency to get the signal we can to a  very good approximation treat the noise curve as flat.

The plan of this paper is as follows. In Sec.\;\ref{sec_lisa_snr}, we present our analytical approximation based on SNR evaluated around the peak frequency, which then evolves over time. We then generalize our formulae to power-law noise curves in Sec.\;\ref{sec_precise} and compare numerically the results of the approximation with those of the true noise curve for LISA and DECIGO in Sec.\;\ref{sec_comparing}. For the latter, we allow for different redshifts (since DECIGO is sensitive to binaries at large redshifts) which simply shifts the overall test to different observed frequencies. We comment on the detection of highly eccentric binaries in Sec.\;\ref{sec_high} and conclude in Sec.\;\ref{sec_conclusions}.

\section{LISA SNR of Eccentric BBHs}
\label{sec_lisa_snr} 

To set the stage, in this section we review the semi-quantitative arguments presented in \cite{Randall:2019znp}, which could help to gain some intuition about the SNR for eccentric events.

We start with the time-domain formula for the SNR \cite{Barack:2003fp}. For a detector like LISA with noise strain $S_N(f)$, the SNR $\varrho$ of eccentric BBHs can be calculated as
\bge
\label{snr_time}
  \varrho^2=2\int_0^{T_O}\di t\,\sum_{n=1}^\infty \FR{ h_n^2(t)}{S_N(f_n(t))},
\ede
where the summation is over all harmonic components of the GWs, $h_n$ is the GW amplitude of $n$'th harmonic component with frequency $f_n=nf_0$, and $f_0$ is the base frequency. $T_O$ is the total observation time. For highly eccentric BBHs we need to sum over a large number of harmonics, which can be numerically challenging. However, the noise strain $S_N(f)$ varies as a power law over much of the range of observable frequencies (except at the turnaround from one power law to another) and this observation can help to simplify the harmonic summation. In fact, it turns out to be a good approximation to pull $S_N$ out of the summation, essentially assuming constant noise.

We then find that the signal would be proportional to $\sum h_n^2=\la h^2\ra\equiv h_c^2$, which is simply the amplitude of GW radiation averaged over one orbit. Having such a formula allows for a much better understanding of eccentricity dependence of the signal to noise.
To see how this quantity depends on eccentricity, first consider the large $e$ limit where we keep track of 
$\ep=1-e^2$ factors. Then the GW amplitude is proportional to $\ddot M$ with $M\propto (a \ep)^2$, the mass quadrupole of the binary, and where $a$ is the semi-major axis of the binary. (See the appendix for more details.) To take the time derivative, we use the fact that $\di/\di t=\dot\psi(\di/\di\psi)$ where $\psi$ is the true anomaly  of the binary orbit on which the mass quadrupole has sinusoidal dependence -- and $\dot\psi\propto (a\ep)^{-3/2}$. Therefore,
\bge
  \la h^2\ra \propto \la \ddot M^2\ra\propto \omega_0\int_0^{2\pi}\di\psi\,\dot\psi^{-1}\ddot M^2\propto a^{-2}\ep^{-1/2}.
\ede
Notice that the four time derivatives from $\ddot M^2$ introduce four factors of $\psidot$ and there is one inverse $\psidot$  from converting the time integral into an integral over the phase. So there are 3 factors of $\dot\psi\propto (a\ep)^{-9/2}$ in total. In addition, the prefactor $\omega_0 \propto a^{-3/2}$. So we have the net result $(a\ep)^4\cdot(a\ep)^{-9/2}\cdot a^{-3/2}\sim a^{-2}\ep^{-1/2}$.

Now we rewrite $\varrho$ in terms of $f_p$ and $\ep$, so we use the formula for $f_p$ derived in \cite{Wen:2002km}, 
\bge
  \label{fp}
  f_p=\FR{\sqrt{Gm}(1+e )^{\ga}}{\pi (a \ep )^{3/2}}, ~~~~\ga=1.1954,
\ede
where $m$ is the total mass of the binary. Below we will also use $n_p\equiv f_p/f_0$. So, using $f_p\propto (a\ep)^{-3/2}$, we have $\la h^2\ra\propto f_p^{4/3}\ep^{3/2}$. There we see if we treat the SNR as a function of $f_p$ and $e$, that the SNR scales with $\ep$ like $\ep^{3/4}$.

\begin{figure}[tbph]
\centering
  \includegraphics[height=0.34\textwidth]{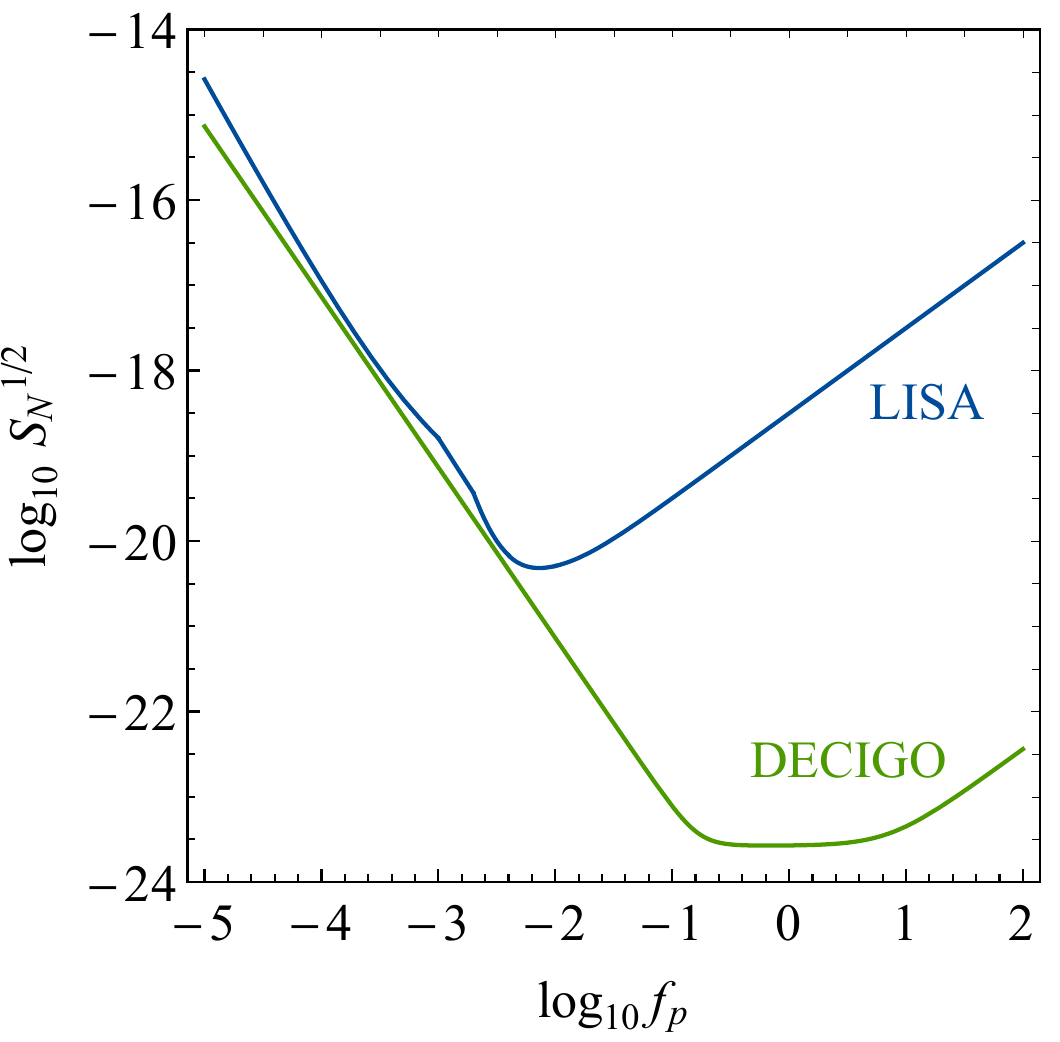} \hspace{1.65cm}
  \includegraphics[height=0.35\textwidth]{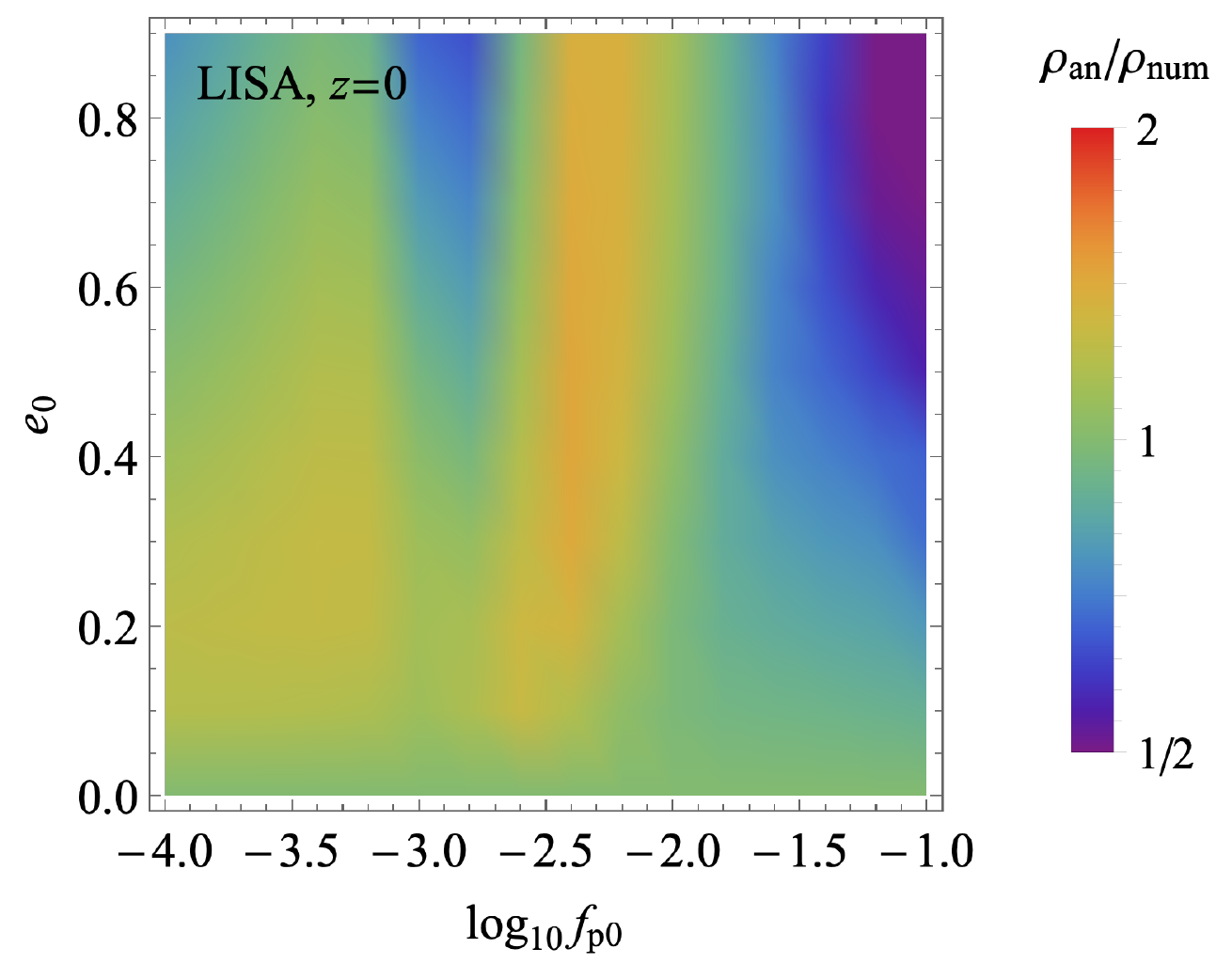}\\[3mm]
  \includegraphics[height=0.35\textwidth]{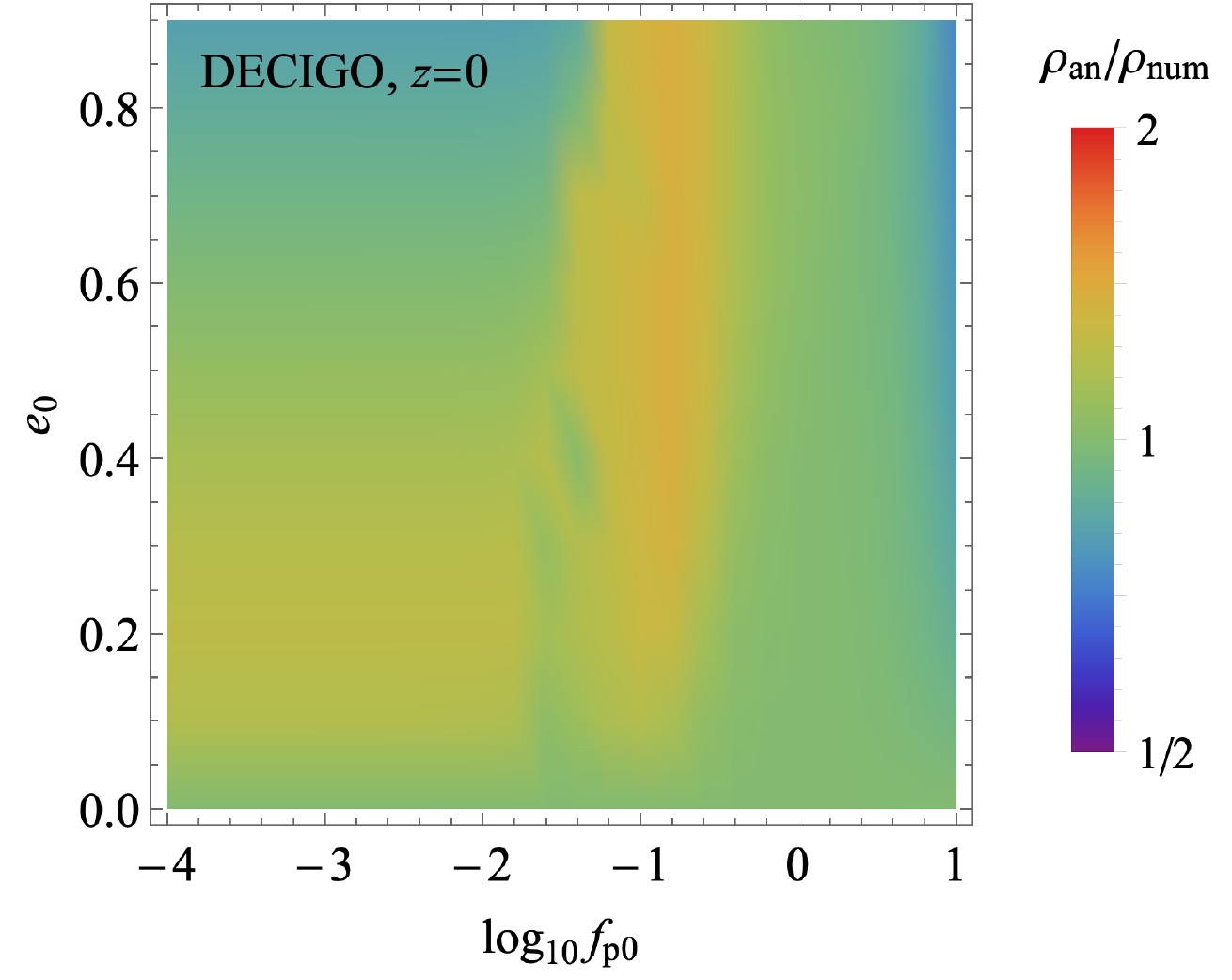}~~
  \includegraphics[height=0.35\textwidth]{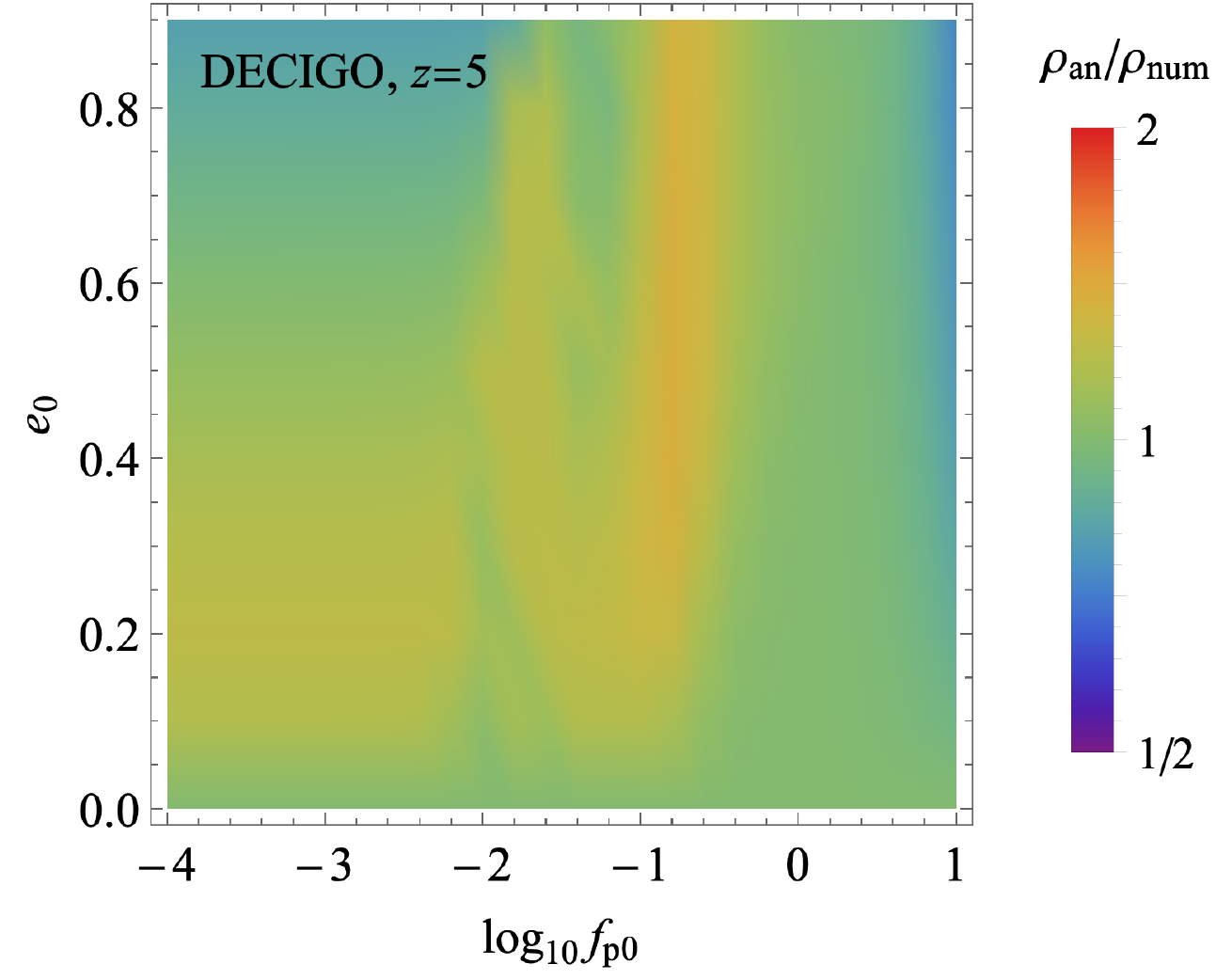} 
  \caption{The SNR $\varrho_\text{an}$ calculated from (\ref{snrsimp}) divided by the SNR $\varrho_\text{num}$ calculated by harmonic summation, plotted as function of initial peak frequency $f_0$ and eccentricity $e_0$. For this calculation we assume a binary with $m_1=m_2=30M_\odot$ and four years of observation time.  UPPER LEFT: The noise curves of LISA \cite{Klein:2015hvg} and DECIGO \cite{Yagi:2011wg} used for this computation. UPPER RIGHT: $\varrho_\text{an}/\varrho_\text{num}$ for LISA at $z=0$. LOWER:  $\varrho_\text{an}/\varrho_\text{num}$ for DECIGO for events that occur at $z=0$ (LOWER LEFT) and $z=5$ (LOWER RIGHT).}
  \label{fig_snr_const}
\end{figure}

Note that this gives us only an approximate formula for the dependence on $1-e$ as $e\to 1$. As in \cite{Randall:2019znp}, to allow for all $0\leq e<1$, we treat the dependence on $1+e$ as an unknown to be determined numerically.  As was shown  in \cite{Randall:2019znp},  a simple formula turns out to well approximate the sum over harmonics for any value of eccentricity $0\leq e<1$:  
\bge
  h_c^2(f_p,e)=h_c^2(f_p,e=0)\cdot(1-e)^{3/2}.
\ede 
Therefore, for binaries with little chirping during the whole observation time meaning 
that $f_p$ and $e$ are relatively constant, we have,
\bge
\label{snrnonchirp}
  \varrho(f_p,e)=\varrho(f_p,e=0)\cdot(1-e)^{3/4}.
\ede 
For chirping binaries, we can extend to the following generalized expression,
\begin{align}
  \label{snrsimp}
  \varrho^2(f_p,e)=4\int\di t\,\FR{h_c^2(f_p(t),e=0)}{S_N(f_p(t))}\big[1-e(t)\big]^{3/2}.
\end{align}
Here $f_p(t)$ and $e(t)$ should be calculated using Peters's equations (\ref{dadt}) and (\ref{dedt}), as well as (\ref{fp}). In this way we can avoid the summation over GW harmonics.  We check the accuracy of our approximation at LISA and DECIGO (for two different values of redshift) in Fig.\;\ref{fig_snr_const}, which shows a comparison of SNR between our formula and an explicit numerical computation summing over harmonics applicable to events at LISA and DECIGO for frequencies up to 0.1Hz and 10Hz respectively.
The inclusion of the redshift affects both the original formula (\ref{snr_time}) and the approximation (\ref{snrsimp}) in the same way (as expected). The only difference between the two lower panels of Fig.\;\ref{fig_snr_const} is the shift in the initial frequency of the events as $f_{p0}/(1+z)$. We include this as a consistency check of our result.

Eq.\ (\ref{snrnonchirp}) shows that a nonzero eccentricity will decrease the SNR for a binary with fixed peak frequency. One can again understand this suppression of SNR by noting that the eccentricity lengthens the lifetime $\tau$ for fixed $f_p$ and thus suppresses the radiation power $\dot E\sim \tau^{-1}$.

 A constant noise strain turns out to be a reasonably good assumption for $e<0.9$ as we can see in Fig.\;\ref{fig_snr_const}.  As we will now see, the  correction comes from deviations from a constant noise curve over frequency. It is nonetheless remarkable that so long as we work in terms of the measurable signal concentrated at and near the peak frequency, we can treat the noise curve as constant, even with significant power law dependence. The result is accurate to at least within a factor of two for eccentricities less than 0.9 over the entire plotted parameter range.  Beyond this, for large eccentricity $e\sim 0.9$, the analytical approximation and the precise formula agree less well, but still within an order of magnitude (as can be seen in the upper right panel of Fig.\;\ref{fig_snr_power}).

 We see that our approximation of a constant noise curve works well in the falling part of the noise curve and the turnaround point.
 In the next section, we derive a formula that is even more accurate on  the falling part and improves the rising part of the noise curve as well.

\section{Precise SNR Formulae for Eccentric Binaries with Power-Law Noise}
\label{sec_precise}

The analysis of \cite{Randall:2019znp} reviewed above assumed a constant noise curve over frequency. In reality, the noise curves for the two lower frequency GW detectors we are considering are projected to have a falling power law dependence ($f^{-4}$ until frequencies $f=5$mHz at LISA and 0.2Hz at DECIGO), where they turn around to follow an $f^2$ dependence. In light of this, assuming a constant noise curve might seem to be a terrible approximation. In this section we will show that the additional dependence is well under control at fixed peak frequency.

\begin{figure}[tbph]
\centering
    \includegraphics[height=0.35\textwidth]{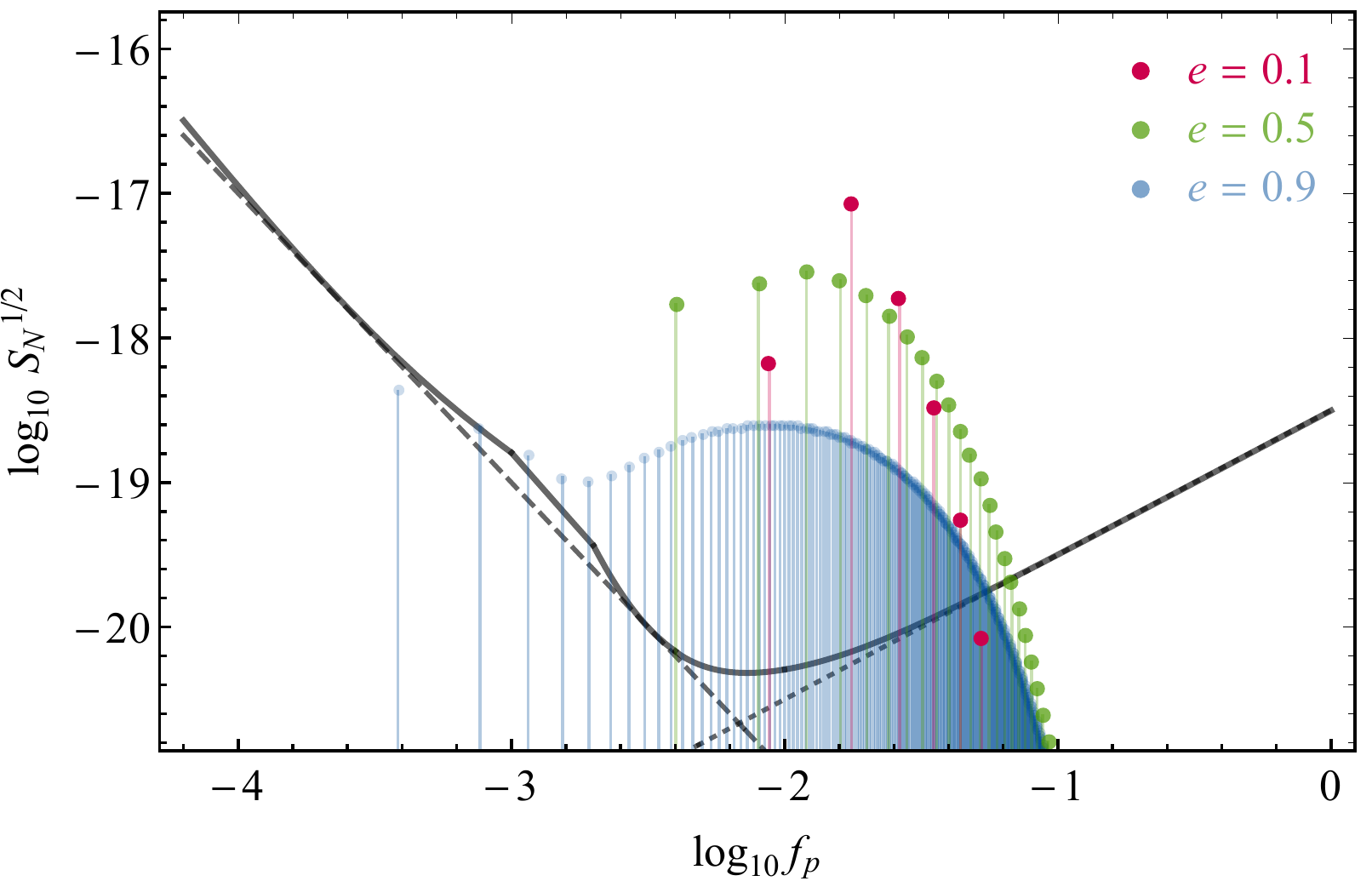}
    \caption{The spectrum of non-chirping eccentric binaries with $f_p=0.01$Hz. The three sets of dots (in magenta, green, and blue) correspond to $e=(0.1,0.5,0.9)$, respectively, with all other parameters are fixed. In the same plot we show the LISA noise $S_N^{1/2}(f)$ in N2A5 configuration (solid black curve). We also draw two curves representing power-law noises with $f^{-4}$ (dashed line) and $f^{2}$ (dotted line). }
    \label{spectrum}
\end{figure}

 In Fig.\;\ref{spectrum} we show the (unnormalized) spectrum of eccentric binaries with $f_p=0.01$Hz and $e=(0.1,0.5,0.9)$, along with the LISA noise curve. As $e$ gets larger, one can clearly observe the broadening of the spectrum due to an increasing number of harmonics. We see that at larger eccentricity, the spectrum develops a secondary peak at the lower frequency end, essentially because of the $1/n$ factor enhancement associated with the amplitude (not the power).

The key insight of the analysis below is that the power-law dependences in the noise is mathematically equivalent to taking time derivatives of the GW amplitudes. We will show that including those additional power dependencies does not heavily affect our previous result for decreasing noise $S_N(f)\sim f^{-4}$ but does bring discrepancies for rising noise $S_N(f)\sim f^2$, allowing us to correct the formula in this case as well. 

We begin with the expression of the SNR in the time domain again, but this time without the harmonic decomposition,
\bge
  \varrho^2=2\int_0^{T_O}\di t\,\FR{1}{S_N(f(t))}\la h_{ij}^\text{TT}(t)h_{ij}^\text{TT}(t)\ra.
\ede
The average $\la\cdots\ra$ in this expression consists of both the angular average and the time average over one orbital motion. The angular average is due to our ignorance of the direction of the orbital plane, and the time average is to make sense of the GW amplitude $h_c$. By the  standard procedure, this average can be related to an average over the  mass quadrupole in the following way. (See the appendix for more details.)
\begin{align}
\label{hc}
  h_c^2(t)\equiv\la h_{ij}^\text{TT}(t)h_{ij}^\text{TT}(t)\ra
  =&~\FR{4G^2}{d^2c^8}\FR{\omega_0}{2\pi}\int_0^{P}\di t \int\FR{\di\Omega}{4\pi}\,\Lambda_{ij,k\ell}\ddot Q_{ij}(t)\ddot Q_{k\ell}(t)\n\\
  =&~\FR{4G^2}{d^2c^8}\FR{\omega_0}{2\pi}\int_0^{2\pi}\di\psi\,\dot\psi^{-1}\FR{2}{5}\Big[\ddot M_{11}^2+\ddot M_{22}^2+2\ddot M_{12}^2-\FR{1}{3}\big(\ddot M_{11}+\ddot M_{22}\big)^2\Big].
\end{align}
Here $d$ is the distance from the binary to the detector (neglecting the redshift), $\omega_0=\sqrt{Gm/a^3}$ is the orbital frequency and $P=2\pi/\omega_0$ is the orbital period. The tensor $\Lambda_{ij,k\ell}$ is the projector onto transverse  and traceless components, $M_{ij}=\mu x_i x_j$ is the second moment of the mass, and $Q_{ij}=M_{ij}-\de_{ij}M_{kk}/3$ is the mass quadrupole.

To compute the SNR, we now decompose the above expression into harmonics, so that the SNR becomes
\bge
  \varrho^2=2\int_0^{T_O}\di t\sum_{n=1}^\infty \FR{1}{S_N(f_n(t))}h_{c,n}^2(t),
\ede
where $h_{c,n}^2(t)$ denotes the $n$'th harmonic component of $h_c^2(t)$. In general, it is the harmonic sum in this expression that makes the computation complicated. This is particularly true for eccentric binaries, because their higher harmonic amplitudes are expressed in terms of the Bessel function $J_n$ of order $n$, whose numerical evaluation can be rather slow. 

However, the computation can be greatly simplified if $S_N(f_n(t))$ is an even integer power of the frequency, as we now show. For a space based detector such as LISA or DECIGO, we recognize three distinct regions in the noise spectrum $S_N(f)$. First, around the minimum, the noise curve $S_N(f)$ can be approximated by a constant up to second order effects. Second, at the low frequency part, the noise curve scales as $S_N(f)\sim f^{-4}$. \footnote{At even lower frequency, e.g., $f<10^{-4}$Hz in LISA, the noise spectrum scales as $S_N(f)\sim f^{-5}$. Our method does not directly generalize to odd powers so does not cover this region. But the noise here is already high enough that this  region is irrelevant to observing stellar-mass binaries.} Third, in the high frequency region, $S_N(f)\sim f^2$. Therefore, we shall consider these three cases in turn.  

\paragraph{Constant noise, $S_N(f)\sim f^0$.} In this case, we can take the noise strain $S_N(f)$ out of the summation. Then we only need to compute $\sum_n h_{c,n}^2(t)=h_{c}^2(t)$, which is nothing but the quantity in (\ref{hc}). Using the explicit expression for $M_{ij}$ and $\dot\psi$, we find
\begin{align}
  h_c^2(t)=&~\FR{64(Gm_c)^{10/3}(\pi f_p)^{4/3}}{5c^8d^2}\mathcal{I}_0(e),\\
  \mathcal{I}_0(e)\equiv &~(1-e^2)^{3/2}(1+e)^{-4\ga/3}\bigg(1+\FR{1-\sqrt{1-e^2}}{3}\bigg),
\end{align}
where $\ga\simeq 1.1954$ is the exponent that appears in the formula for the peak frequency. So we find
\bge
\label{snr_const}
  \FR{\varrho(f_p,e)}{\varrho(f_p,e=0)}\Big|_{S_N=\text{const}}=\mathcal{I}_0(e).
\ede
 
\begin{figure}[tbph]
\centering
  \includegraphics[height=0.28\textwidth]{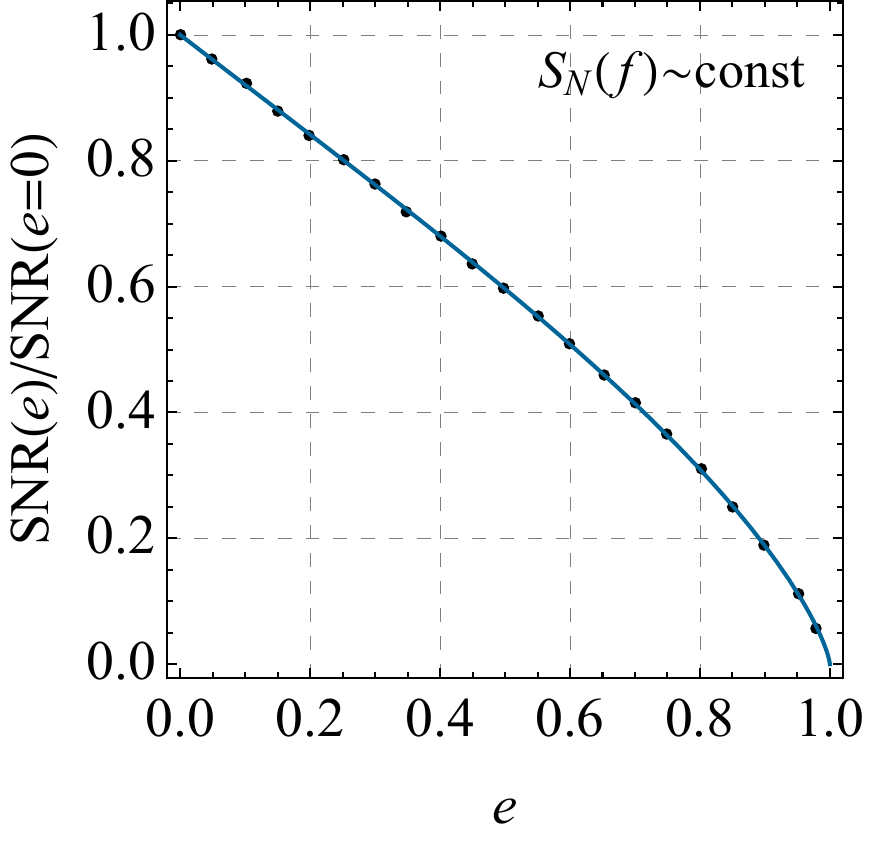}
  \includegraphics[height=0.28\textwidth]{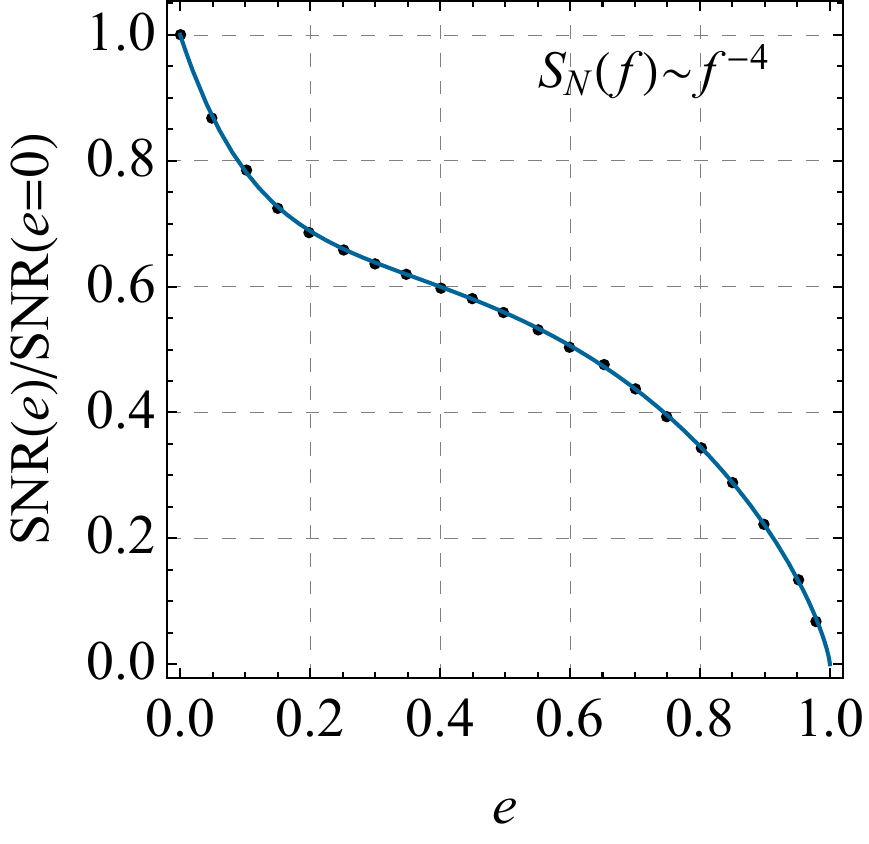}
  \includegraphics[height=0.28\textwidth]{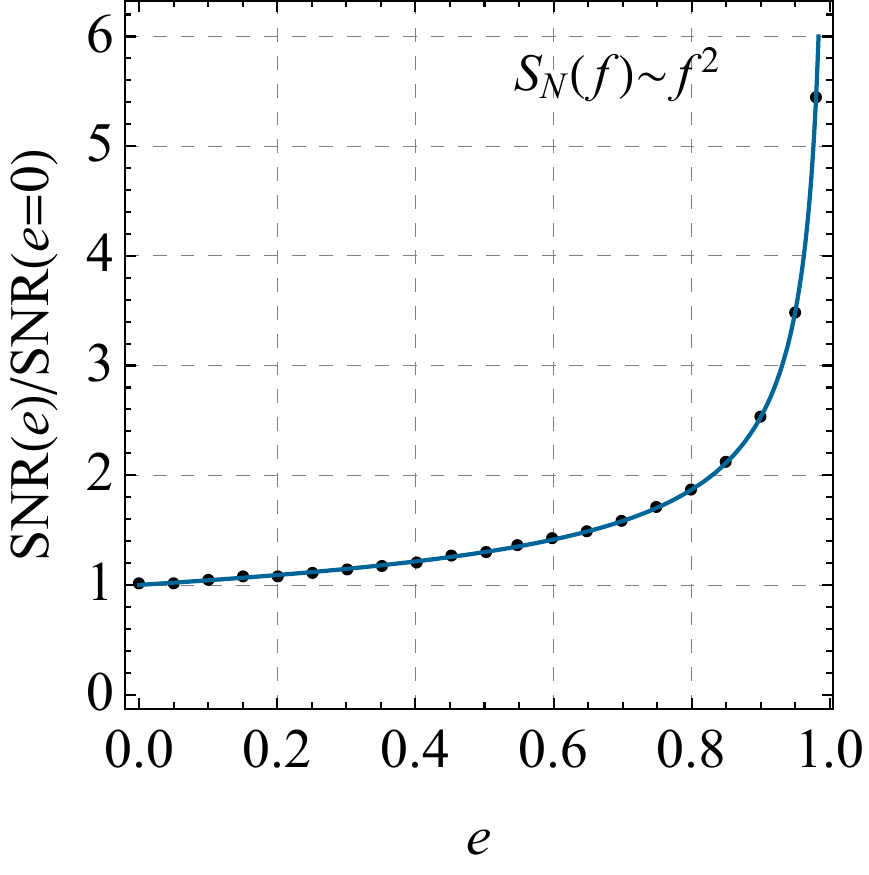}
  \caption{The SNR of eccentric binaries with power-law noises, normalized by the SNR of a circular binary with the same peak frequency. The three panels from left to right correspond to $S_N(f)\sim(f^0,f^{-4},f^2)$,respectively. The black dots are results of harmonic summations, and the blue curves are computed using (\ref{snr_const}), (\ref{snr_m4}), and (\ref{snr_2}), respectively. }
  \label{fig_snr_e}
\end{figure}

\paragraph{Decreasing noise, $S_N(f)\propto f^{-4}$.} In this case we can no longer take $S_N(f)$ out of the summation. But we observe that $S_N(f_n)$, where $f_n=n f_p/n_p$, is $S_N(f_n)=S_N(f_p)(n/n_p)^{-4}$, so $S_N(f_p)$ can be taken outside the summation. Here $n_p$ is the harmonic number of the peak harmonic. So the SNR becomes
\bge
  \varrho^2=2\int_0^{T_O}\di t\,\FR{1}{S_N(f_p)}\sum_{n=1}^{\infty}\FR{n^4}{n_p^4}h_{c,n}^2(t).
\ede
To complete the summation here, we observe that $h_{n,ij}^\text{TT}\propto \cos (n\omega_0 t)$ or $\sin (n\omega_0 t)$. So,
\bge
  \ddot h_{n,ij}^\text{TT}=-n^2\omega_0^2 h_{n,ij}^\text{TT}=-(2\pi f_p)^2\FR{n^2}{n_p^2}h_{n,ij}^\text{TT}.
\ede
Therefore,
\begin{align}
  \sum_{n=1}^{\infty}\FR{n^4}{n_p^4}h_{c,n}^2(t)=&~\FR{1}{(2\pi f_p)^4}\la\ddot h_{ij}^\text{TT}(t)\ddot h_{ij}^\text{TT}(t)\ra\n\\
  =&~\FR{1}{(2\pi f_p)^4}\FR{4G^2}{d^2c^8}\FR{\omega_0}{2\pi}\int_0^{2\pi}\di\psi\,\dot\psi^{-1}\FR{2}{5}\Big[\ddddot {M\hspace{0pt}}_{11}^2+\ddddot {M\hspace{0pt}}_{22}^2+2\ddddot {M\hspace{0pt}}_{12}^2-\FR{1}{3}\big(\ddddot {M\hspace{0pt}}_{11}+\ddddot {M\hspace{0pt}}_{22}\big)^2\Big].
\end{align}
Finishing the integral, we get
\begin{align}
  \sum_{n=1}^{\infty}\FR{n^4}{n_p^4}h_{c,n}^2=&~\FR{64(Gm_c)^{10/3}(\pi f_p)^{4/3}}{5c^8d^2}\mathcal{I}_{-4}(e),\\
  \mathcal{I}_{-4}(e)\equiv &~(1-e^2)^{3/2}(1+e)^{-16\ga/3}\Big(1+\FR{85}{6}e^2+\FR{5171}{192}e^4+\FR{1751}{192}e^6+\FR{297}{1024}e^8\Big).
\end{align}
So we get
\bge
\label{snr_m4}
  \FR{\varrho(f_p,e)}{\varrho(f_p,e=0)}\Big|_{S_N\propto f^{-4}}=\mathcal{I}_{-4}(e).
\ede
\paragraph{Increasing noise, $S_N(f)\propto f^2$.} Using the reasoning similar to that above, we see that this time we need to compute
\begin{align}
  \sum_{n=1}^{\infty}\FR{n_p^2}{n^2}h_{c,n}^2=&~(2\pi f_p)^2\FR{4G^2}{d^2c^8}\FR{\omega_0}{2\pi}\int_0^{2\pi}\di\psi\,\dot\psi^{-1}\FR{2}{5}\Big[\dot {M\hspace{0pt}}_{11}^2+\dot {M\hspace{0pt}}_{22}^2+2\dot {M\hspace{0pt}}_{12}^2-\FR{1}{3}\big(\dot {M\hspace{0pt}}_{11}+\dot {M\hspace{0pt}}_{22}\big)^2\Big]\n\\
  =&~\FR{64(Gm_c)^{10/3}(\pi f_p)^{4/3}}{5c^8d^2}\mathcal{I}_{2}(e),\\
  \mathcal{I}_2(e)\equiv&~\FR{1}{1-e^2}(1+e)^{2\ga/3}\Big(1-\FR{1}{3}e^2\Big).
\end{align}
Therefore, for this rising noise we have
\bge
\label{snr_2}
  \FR{\varrho(f_p,e)}{\varrho(f_p,e=0)}\Big|_{S_N\propto f^{2}}=\mathcal{I}_{2}(e).
\ede
In Fig.\;\ref{fig_snr_e} we show the three analytical formulae for SNR, (\ref{snr_const}), (\ref{snr_m4}), and (\ref{snr_2}), corresponding to three  power laws $S_N\sim (f^0,f^{-4},f^2)$ of the noise curve. We compare these results with explicit harmonic summation and find perfect agreement as we should since the formula is exact in this case. This figure also shows clearly why the constant-noise assumption works well even when when the  noise curve scales as $f^{-4}$: This is essentially because $\mathcal{I}_0$ and $\mathcal{I}_{-4}$ are numerically very similar. On the other hand, the SNR scales with $e$ in the opposite direction when $S\sim f^2$, showing that the constant-noise assumption will not work well for the rising part of the noise curve, and an improved formula (\ref{snr_2}) was called for. We will confirm these observations in the next section by an explicit numerical check.

We note that the LISA noise strain $S_N(f)$ increases following a power law at both sides of the LISA frequency band, namely $S_N(f)\propto f^2$ on the high frequency side ($f>0.1$Hz) and $S_N(f)\propto f^{-4}$ at low frequencies ($f<1$mHz). We can now substitute the above power law into (\ref{snr_time}) and redo the summation over harmonics. The $n$th harmonic introduces additional factors of powers of $n$ for the associated time derivative and we are not left with the simple sum we had before. 
  However, we saw that this $n$-dependence can be replaced by a time derivative. For $a \epsilon$ constant as it is for fixed peak frequency, any  additional eccentricity dependence introduced cancels (though the overall factor can change). This is a remarkable result.

\section{Comparing with Numerical Results}
\label{sec_comparing}

In this section we compare the SNR calculated using our approximating formulae with the numerical results obtained by directly summing over harmonics. We do this comparison for different choices of parameters and instruments with increasing generality. 

\begin{figure}[tbph]
\centering
  \hspace{-8.5mm}
  \includegraphics[height=0.33\textwidth]{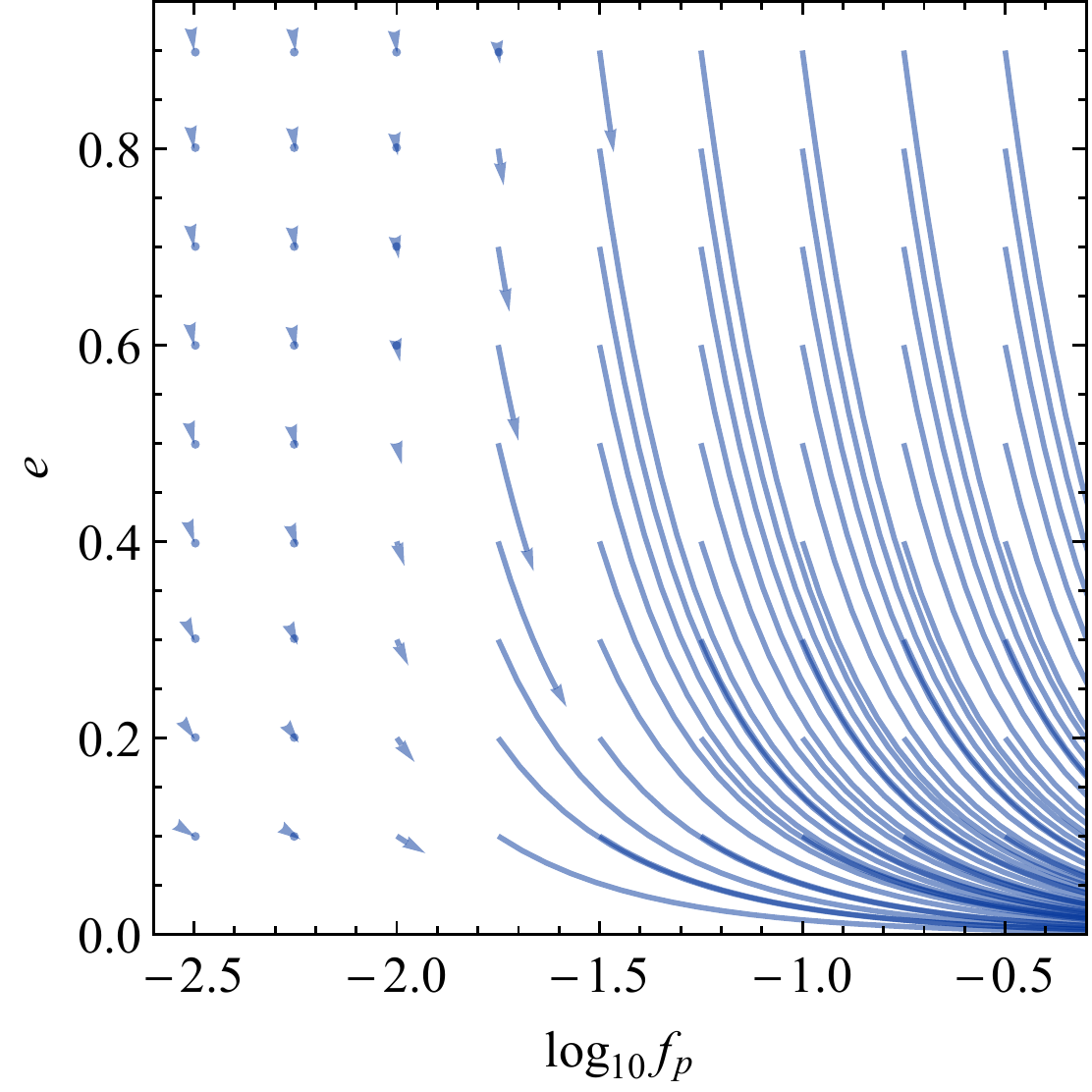}~~~~~~
  \includegraphics[height=0.28\textwidth]{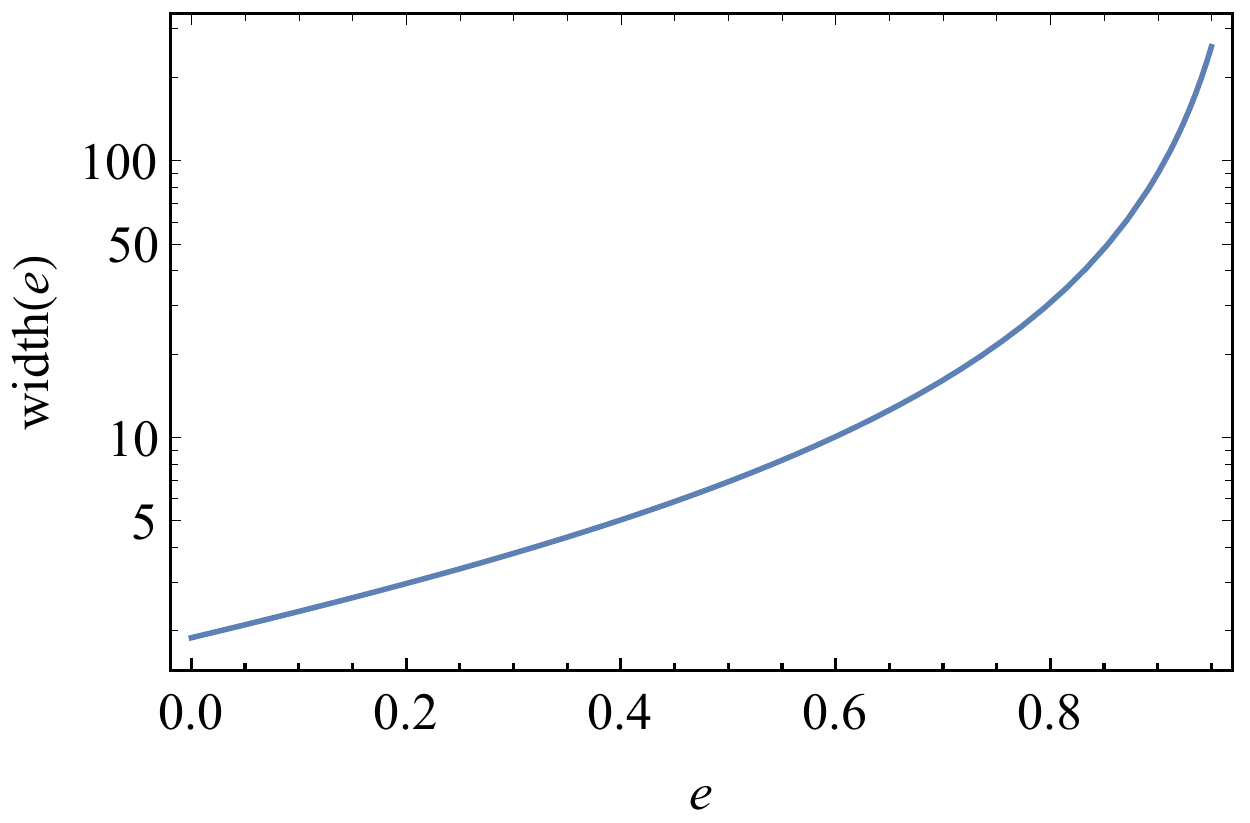} 
  \caption{(Left) The evolution of binaries over fur years with various initial peak frequencies $f_p$ and eccentricities $e$. (Right) The width of the GW spectrum of a binary with eccentricity $e$.}
  \label{fig_chirp}
\end{figure}

First,  we consider a chirping binary of black hole with $m_1=m_2=30M_\odot$ and compute its SNR collected in LISA. For this computation we assume N2A5 configuration and four years of total observation time, and assume the events arise at low redshift.\footnote{We generalize our expression to higher redshift in a forthcoming publication \cite{Randall:future}.} We calculate the ratio $\varrho_\text{an}/\varrho_\text{num}$ where $\varrho_\text{an}$ is calculated using our analytical formulae, and $\varrho_\text{num}$ is from the harmonic summation. During the four-year observation, the binary will chirp ($\dot f_{p}>0$) and circularize ($\dot e<0$ if $e\neq 0$). The rates of chirping and circularization depend on $f_p$ and $e$. To get a better intuition about this, we show in Fig.\;\ref{fig_chirp} the evolution of binaries in the $(f_p,e)$ plane over four years. To remind the reader that the actual spectra of these eccentric binaries have finite width, we show in the right panel of Fig.\;\ref{fig_chirp} the ``width'' of the spectrum of an eccentric binary. Here the ``width'' is defined to be the higher harmonic number at which the radiation power is half of its peak value at $f_p$. 

 We note that on the decreasing part of the noise curve, the frequency changes very little due to chirping. However, at higher frequencies and, in particular, near the dip of the LISA noise curve, the chirping changes the frequency significantly. This will be important below, where the approximation works very well apart from events with high eccentricity at frequencies near but higher than the frequency at  the dip of the noise curve.

\begin{figure}[tbph]
\centering
  \includegraphics[height=0.35\textwidth]{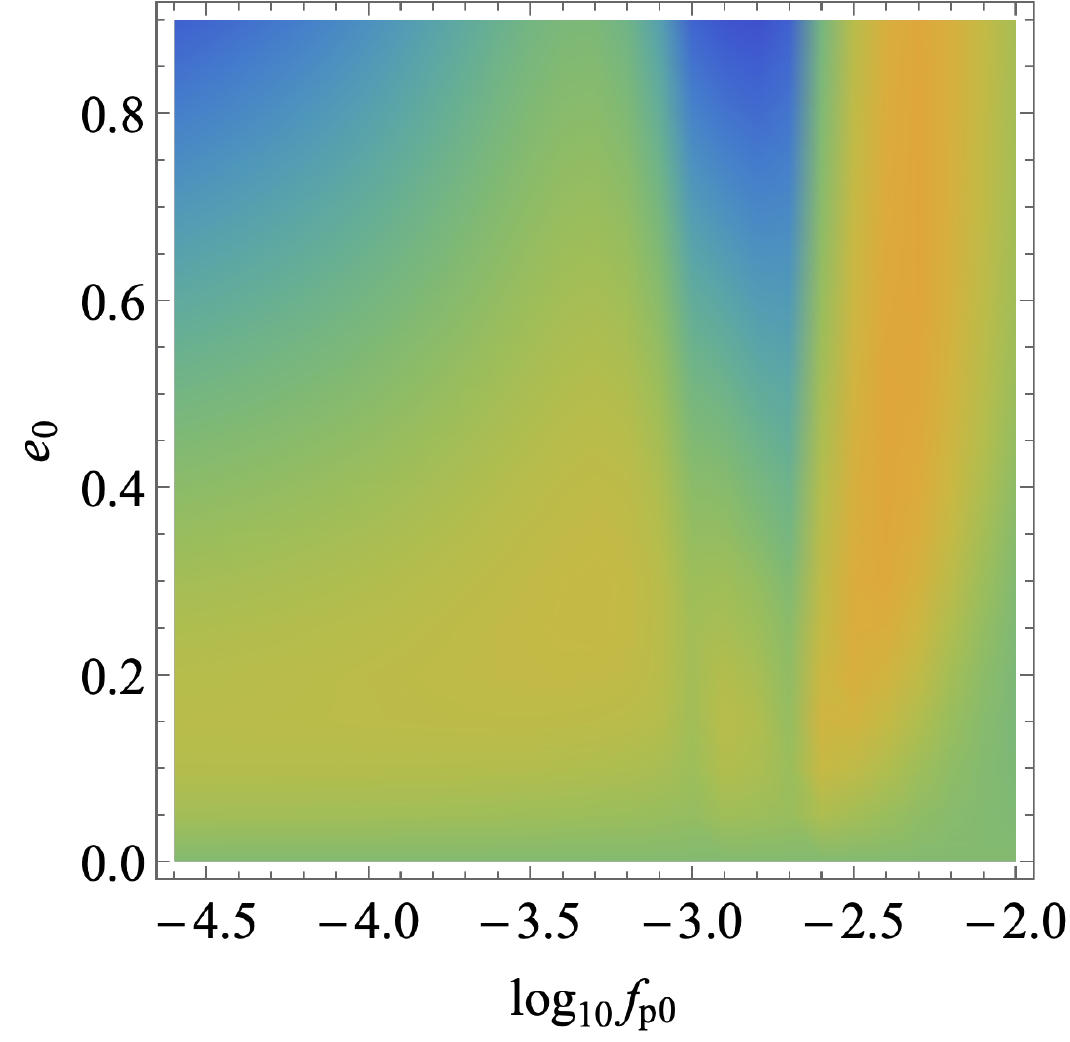}
  \includegraphics[height=0.35\textwidth]{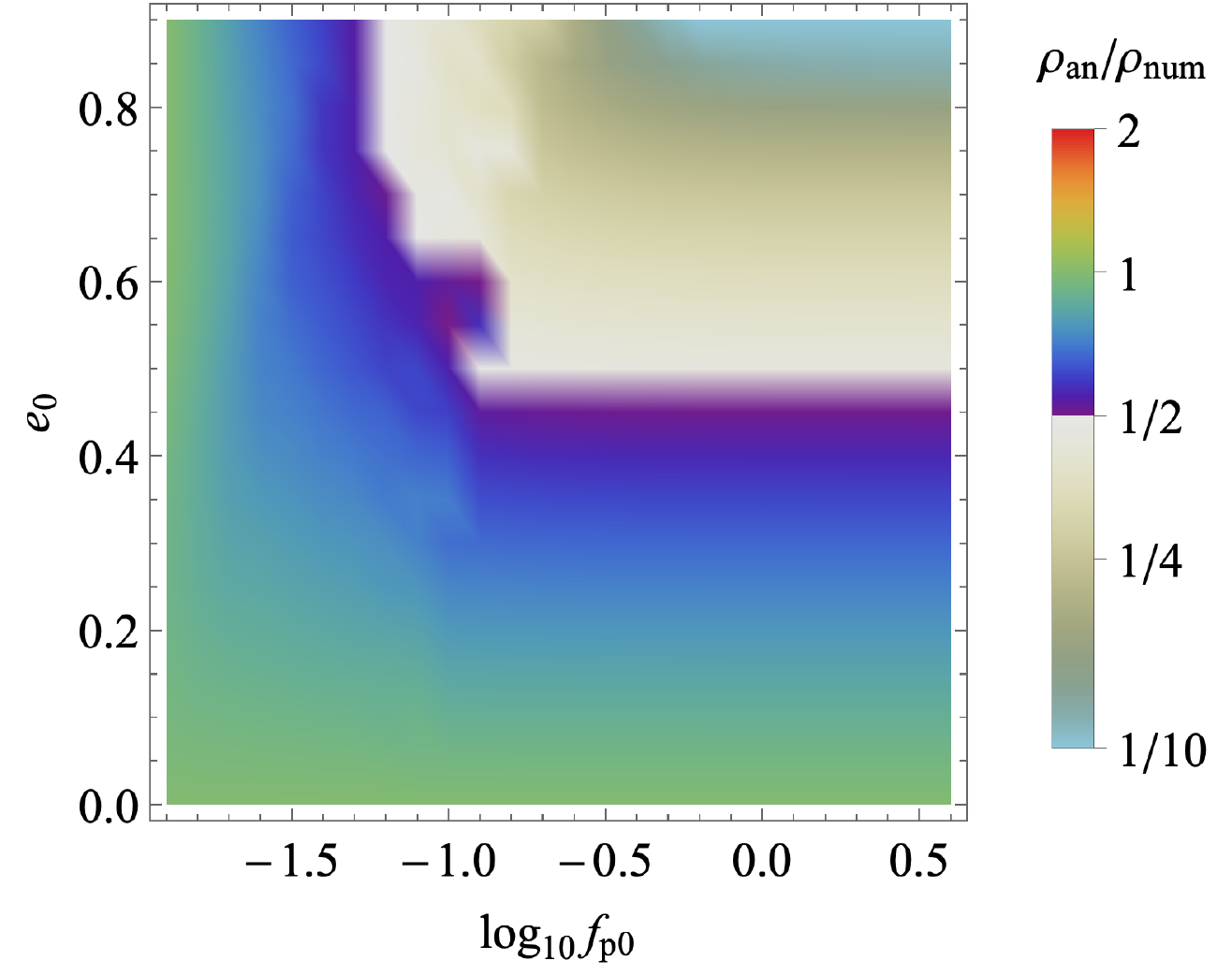}\\
  \includegraphics[height=0.35\textwidth]{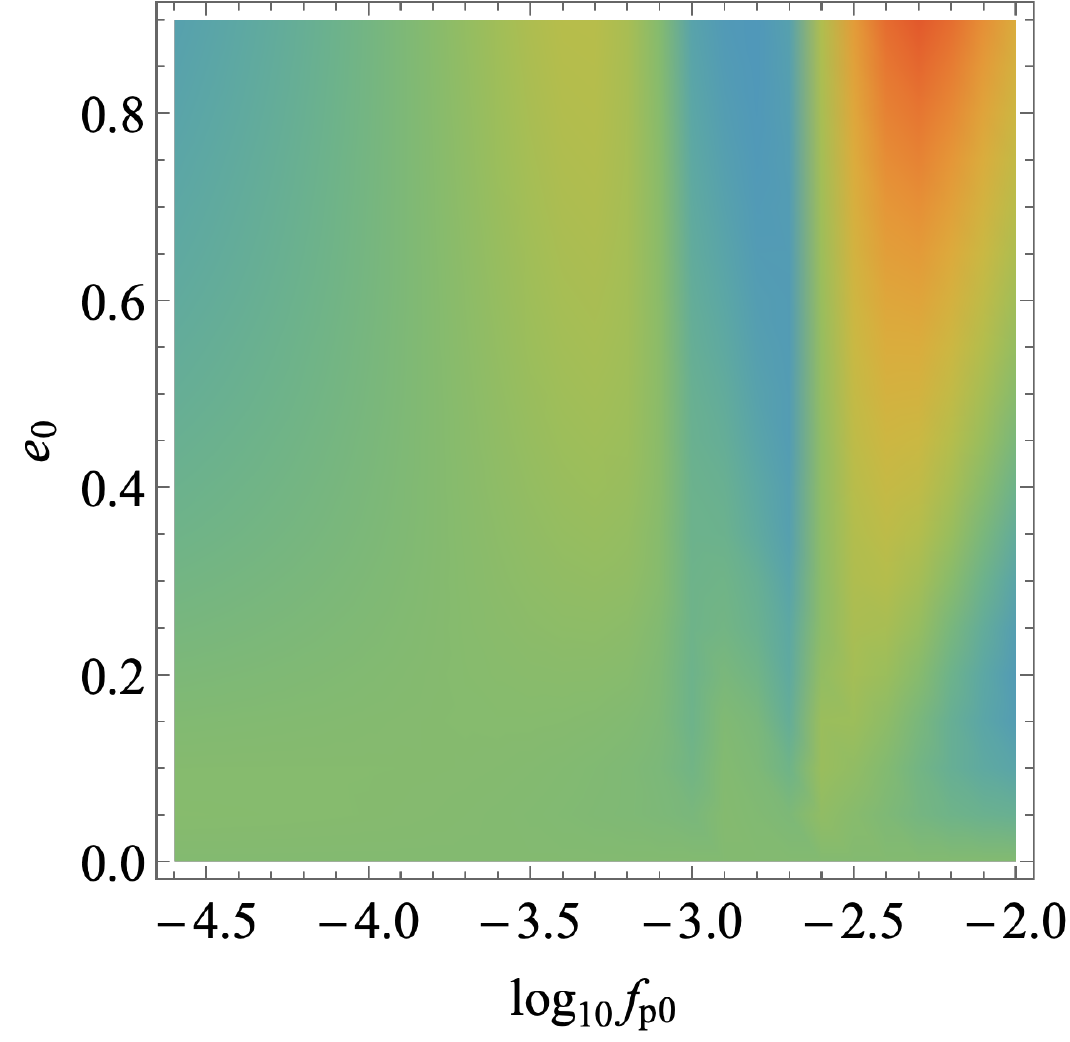}
  \includegraphics[height=0.35\textwidth]{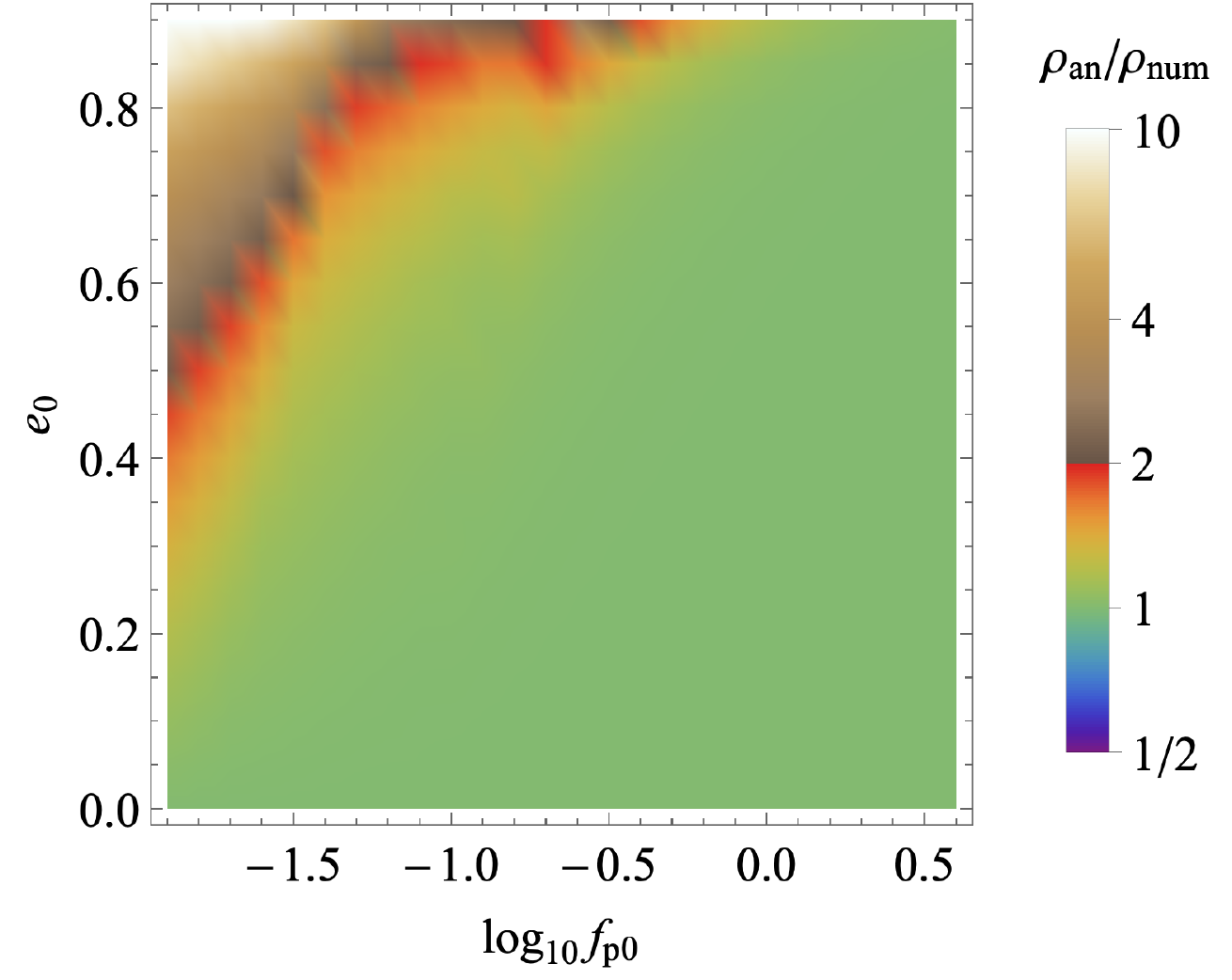}
  \caption{The SNR calculated from analytical formulae in Sec.\;\ref{sec_precise} divided by the SNR calculated by harmonic summation, plotted as function of the peak frequency $f_0$ and the eccentricity $e_0$  at the initial time of observation. For this calculation we assume a binary with $m_1=m_2=30M_\odot$ at $z=0$ and four years of observation time. The two upper panels show the result of constant-noise formula (\ref{snr_const}) divided into two panels corresponding to the approximation to the noise curve used in the lower panels. The lower-left panel and lower-right panel show the results of decreasing-noise formula (\ref{snr_m4}) and increasing-noise formula (\ref{snr_2}), respectively. Notice the coordinate range here extends further than in Fig.\;\ref{fig_snr_const}. }
  \label{fig_snr_power}
\end{figure}

We consider the three analytical formulae derived in the previous section. In the two upper panels of Fig.\;\ref{fig_snr_power} we show the result for the constant noise formula  (\ref{snr_const}). We split the result into two panels, corresponding to the low-frequency/decreasing-noise and high-frequency/increasing-noise parts of the LISA, respectively. We see that the constant noise formula gives a decent estimate of the SNR for most parameter space. Except for the region with $f_p\gtrsim 0.1$Hz and $e_0\gtrsim 0.5$, the constant-noise formula and the numerical result always agrees within a factor of two. As expected, the constant-noise formula works best around the minimum of the noise curve, corresponding to $f_p\simeq 0.01$. The formula also works relatively well for the decreasing part of the noise curve. This is explained by the fact that the constant-noise formula  and  the decreasing-noise formula have the same limiting behavior when $e\to 1$. The constant-noise formula does not work  so well for high frequencies ($f_p>0.1$Hz) because the noise curve is rising here and the actual SNR has opposite asymptotic behavior as $e\to 1$, as explicitly shown by the increasing-noise formula. Finally, there some features in the region $10^{-3}\text{Hz}<f_p<10^{-2}\text{Hz}$ and this is due to the galactic noise in the noise curve,  which cannot be modeled by a simple power. 

We then compare the decreasing-noise formula (\ref{snr_m4}) with the numerical result in the lower-left panel of Fig.\;\ref{fig_snr_power}. We can see clear improvement compared to the constant-noise formula in the low frequency part when $f_p<10^{-3}$Hz, where the noise curve is well approximated by $f^{-4}$ so our analytical formula becomes asymptotically exact. The lack of agreement compared to the constant-noise curve for higher frequency $f_p\gtrsim 10^{-2.5}$Hz is due to the galactic noise, which makes the power law approximation less accurate in this region.

In the lower-right panel of Fig.\;\ref{fig_snr_power} we show the comparison between the increasing-noise formula (\ref{snr_2}) and the numerical result. This time we see clear improvement at higher frequencies compared with the constant-noise formula. And the quality of the approximation has the opposite asymptotic behavior in the two right panels, as expected. For very large eccentricity, we see that the increasing-noise formula always tends to overestimate the SNR, because this formula assumes an increasing noise curve everywhere and thus would underestimate the noise at sufficiently low frequencies, where  the actual noise curve deviates from the $S_N(f)\sim f^2$ behavior. This overestimated SNR becomes more significant  for higher eccentricity because a more eccentric binaries will radiate a wider spectrum extending to lower frequencies. 

As can be seen from the two right panels of Fig.\;\ref{fig_snr_power} there is a small region ($f_{p0}\sim0.1$Hz, $e_0>0.8$) where the analytical formula does not work as  well, yielding more than a factor of two error. One needs to be careful when applying our formula to compute SNR for binaries in this region. However, here we benefit from the change of frequency due to chirping at high frequency illustrated in Fig.\;\ref{fig_chirp}. An event moves out of the regime where the approximation works less well. Applying similar reasoning to DECIGO, the dip in noise occurs at even higher frequency where the chirping is even more significant so our approximation should apply at least as well in that frequency regime.

\section{High Eccentricity Detection}
\label{sec_high}

We note a particularly interesting case when the eccentric binaries' peak frequency $f_p$ sits at the rising part of the noise curve where $S_N(f)\propto f^2$. As shown above, the SNR diverges for large eccentricity like $\rho^2\sim (1-e^2)^{-1}$ with other parameters fixed. This is due to the broadened GW spectrum at large $e$ so that the spectrum extends to the lower frequency part where the noise is also lower. This suggests an enhanced detectability of highly eccentric binaries at the rising part of the noise curve over the naive expectation.  
This point was often ignored in previous studies, but can be  crucial for highly eccentric binaries formed with frequencies where the detector noise curve is rising. This corresponds to $f_p\gtrsim0.01$ for LISA and $f_p\gtrsim 0.1$ for DECIGO.

It is interesting to check whether we expect it to be  possible  to detect any  eccentric binaries of this category, namely $f_p>0.01$ for LISA (or $f_p>0.1$ for DECIGO) and $e>0.9$. A quick and dirty estimate for the number of observable binaries of this sort can be done by looking at LIGO events with (in principle) detectable nonzero eccentricity, namely, $e>0.01$ at $f_p=10$Hz. It was derived in \cite{Randall:2019znp} that the eccentricity $e$ and the peak frequency $f_p$ for a binary are related by
\begin{align}
    &\FR{f_p}{f_{p*}}=\FR{\mathcal{H}(e)}{\mathcal{H}(e_*)}, 
    &&\mathcal{H}(e)\equiv\FR{(1+e)^\ga}{[(1-e^2)\mathcal{G}(e)]^{3/2}},
    &&\mathcal{G}(e)\equiv\FR{e^{12/19}}{1-e^2}\bigg(1+\FR{121}{304}e^2\bigg)^{870/2299}.
\end{align}
From this we see that any binaries with $e>0.01$ at $f_p=10$Hz would have $e>0.9$ at the rising part of the LISA noise. 

Many dynamical channels predict the existence of such highly eccentric binaries. For example, \cite{Martinez:2020lzt} predicts a local merger rate $0.35\text{Gpc}^{-3}\text{yr}^{-1}$ for the dynamical binaries formed in dense star clusters and $\sim 9\%$ of them would have $e>0.01$ at $10$Hz, corresponding to large eccentricity in the sub-Hertz region. \cite{Martinez:2020lzt} argued that around 80\% percent of the KL binaries and 20\% of non-KL binaries from this channel can have large eccentricity at $f_p\gtrsim 0.01$Hz in LISA.  There are also other channels that can produce similar highly eccentric mergers. In any case, when these mergers appear in the rising part of LISA or DECIGO noise, they can carry significantly large eccentricity, so that our method here could be quite useful.

\section{Conclusions} 
\label{sec_conclusions}

\begin{figure}[tbph]
\centering
  \includegraphics[width=0.8\textwidth]{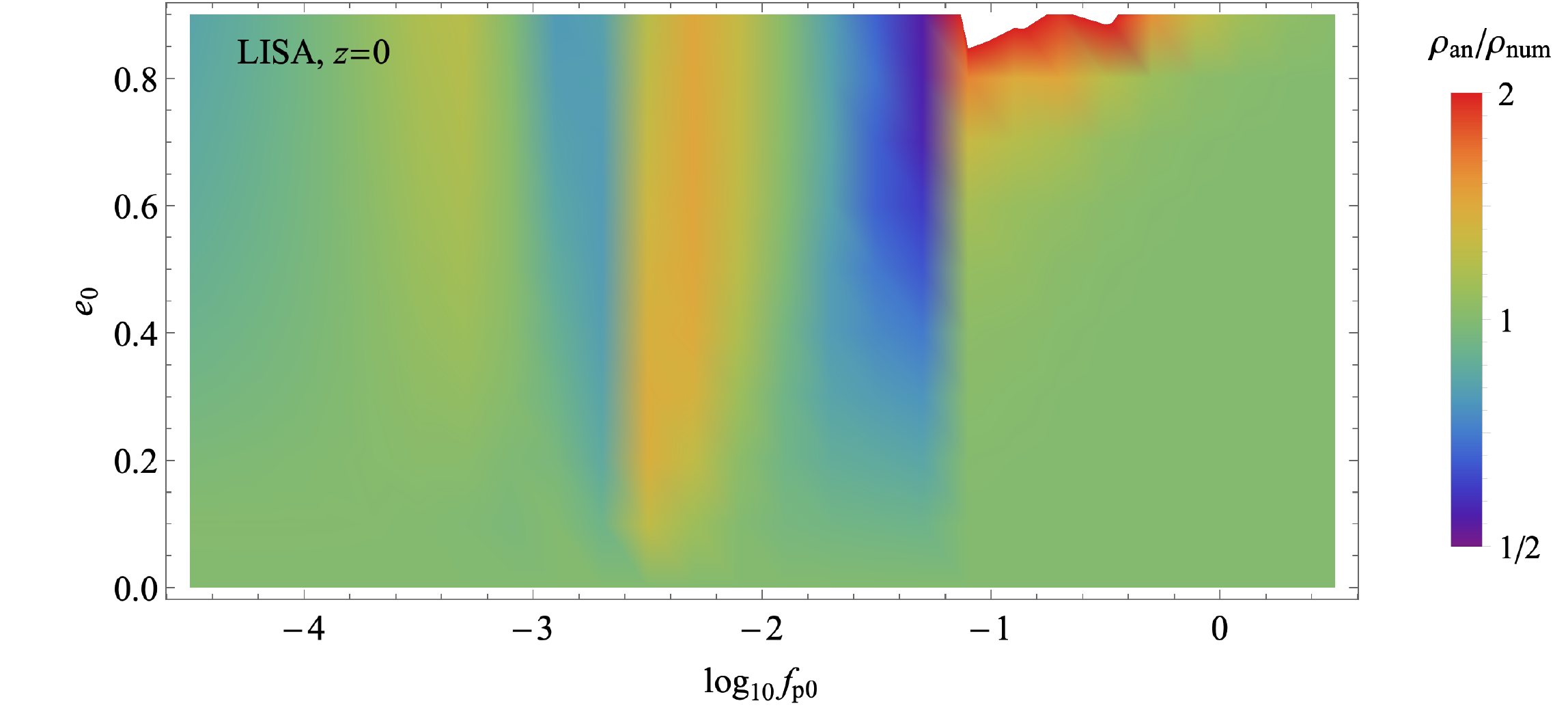}
  \caption{The LISA SNR calculated from analytical formulae in Sec.\;\ref{sec_precise} divided by the SNR calculated by harmonic summation, plotted as function of the peak frequency $f_0$ and the eccentricity $e_0$  at the initial time of observation. For this calculation we assume a binary with $m_1=m_2=30M_\odot$ at $z=0$ and four years of observation time. We use (\ref{snr_m4}), (\ref{snr_const}), and (\ref{snr_2}) for $f_{p0}<10^{-2.6}$Hz, $10^{-2.6}$Hz$< f_{p0}< 10^{-1.3}$Hz, and $f_{p0}>10^{-1.3}$Hz, respectively. }
  \label{fig_fullrange}
\end{figure}

A great deal of work has been devoted to detailed predictions of the wave forms for GW signals and building templates to accommodate them. However, very little attention has been paid to approximations that can be useful when predicting distributions and also when narrowing the parameter space for existing events. We have shown how to readily approximate SNR, even for highly eccentric events, in future lower frequency detectors.  We summarize our result over the entire LISA frequency range in Fig.\;\ref{fig_fullrange}. Furthermore this type of analysis can sometimes highlight interesting features of the proposed detections, such as the anomalously big potential SNR for highly eccentric events. 

In a future work we will follow up our analysis on relating distributions at different frequencies when the events have initially high eccentricity. In particular we can use the method outlined here to estimate the importance of intermediate frequency bands when distinguishing populations.

In this paper our main focus has been on the eccentric stellar-mass binaries. But we emphasize that our method can be directly generalized to any quasi-period sources. Notable examples include intermediate-mass black hole binaries, extreme mass ratio inspirals, and spinning compact objects. We look forward to interesting predictions for those in the future.

\paragraph{Acknowledgment} We thank Karan Jani and Nick DePorzio for useful discussions and Nick Deporzio for helpful comments on the manuscript.  LR was supported by an NSF grant PHY-1620806, the Chan Foundation, a Kavli Foundation grant ``Kavli Dream Team,'' the Simons Fellows Program, 
the Guggenheim Foundation, and a Moore Foundation Fellowship.
ZZX is supported by Tsinghua University Initiative Scientific Research Program.

\begin{appendix}

\section{More details}
The GW amplitude is related to the mass quadrupole $Q_{ij}$ by
\bge
  h_{ij}^\text{TT}=\FR{1}{d}\FR{2G}{c^4}\Lambda_{ij,k\ell}\ddot Q_{k\ell}.
\ede

Here $Q_{ij}=M_{ij}-\de_{ij}M_{kk}/3$ and $M_{ij}=\mu x_i x_j$. The projector $\Lambda_{ij,k\ell}$ projects a tensor to its symmetric, transverse, and traceless part,
\begin{align}
  &\Lambda_{ij,k\ell}(\hat{\mb n})=P_{ik}P_{j\ell}-\FR{1}{2}P_{ij}P_{k\ell}, &&P_{ij}(\hat{\mb n})=\de_{ij}-n_i n_j.
\end{align}
To carry out the angular integral in (\ref{hc}), it is useful to note that 
\bge
  \int\FR{\di\Omega}{4\pi}\Lambda_{ij,k\ell}=\FR{1}{30}(11\de_{ik}\de_{j\ell}-4\de_{ij}\de_{k\ell}+\de_{i\ell}\de_{jk}).
\ede
So,
\bge
  \int\FR{\di\Omega}{4\pi}\Lambda_{ij,k\ell}Q_{ij}Q_{k\ell}=\FR{2}{5}Q_{ij}Q_{ij}.
\ede
(Note that $Q_{ij}$ is traceless.) Rewriting $Q$ in terms of $M$ then gives Eq. (\ref{hc}).

Then we need to know what is $M$ and how to compute its time derivative.

By rotation symmetry we can always bring the elliptical orbit into the $(x,y)$ plane. Then, using the polar coordinates $(r,\psi)$, the second moment of the mass $M_{ij}$ is 
\bge
  M_{ij}=\mu r^2\bgp \cos^2\psi & \cos\psi\sin\psi \\ \cos\psi\sin\psi & \sin^2\psi \edp.
\ede
For an elliptical orbit, $r$ can be rewritten as $r=a(1-e^2)/(1+e\cos\psi)^2$. So we have expressed $M_{ij}$, and thus $h_{ij}$, in terms of $\psi$. It is this phase $\psi$ that has fast periodic motion. So, to take time derivatives on $M_{ij}$, we only need to know how to take time derivative on $\psi$, and this is nothing but Kepler's second law:
\bge
  \dot\psi=\sqrt{\FR{Gm}{a^3(1-e^2)^3}}(1+e\cos\psi)^2.
\ede
The chirping and circularization of an eccentric binary by radiating GWs at quadrupole order are described by Peters's equations \cite{Peters:1964zz},
\begin{align}
\label{dadt}
\dot a=&-\FR{64}{5}\FR{G^3\mu m^2}{c^5a ^3}\FR{1}{(1-e^2)^{7/2}}\bigg(1+\FR{73}{24}e^2+\FR{37}{96}e^4\bigg),\\
\label{dedt}
\dot e =&-\FR{304}{15}\FR{G^3\mu m^2}{c^5a ^4}\FR{e}{(1-e^2)^{5/2}}\bigg(1+\FR{121}{304}e^2\bigg).
\end{align}

\end{appendix}

\providecommand{\href}[2]{#2}\begingroup\raggedright\endgroup


\begin{thebibliography}{10}

\bibitem{LIGOScientific:2018jsj}
{\bfseries LIGO Scientific, Virgo} Collaboration, B.~P. Abbott {\em et~al.},
  ``{Binary Black Hole Population Properties Inferred from the First and Second
  Observing Runs of Advanced LIGO and Advanced Virgo},''
\href{http://arxiv.org/abs/1811.12940}{{\ttfamily arXiv:1811.12940
  [astro-ph.HE]}}.

\bibitem{LIGOScientific:2018mvr}
{\bfseries LIGO Scientific, Virgo} Collaboration, B.~P. Abbott {\em et~al.},
  ``{GWTC-1: A Gravitational-Wave Transient Catalog of Compact Binary Mergers
  Observed by LIGO and Virgo during the First and Second Observing Runs},''
  \href{http://dx.doi.org/10.1103/PhysRevX.9.031040}{{\em Phys. Rev. X}
  {\bfseries 9} no.~3, (2019) 031040},
  \href{http://arxiv.org/abs/1811.12907}{{\ttfamily arXiv:1811.12907
  [astro-ph.HE]}}.

\bibitem{Dominik:2012kk}
M.~Dominik, K.~Belczynski, C.~Fryer, D.~Holz, E.~Berti, T.~Bulik, I.~Mandel,
  and R.~O'Shaughnessy, ``{Double Compact Objects I: The Significance of the
  Common Envelope on Merger Rates},''
  \href{http://dx.doi.org/10.1088/0004-637X/759/1/52}{{\em Astrophys. J.}
  {\bfseries 759} (2012) 52}, \href{http://arxiv.org/abs/1202.4901}{{\ttfamily
  arXiv:1202.4901 [astro-ph.HE]}}.

\bibitem{Dominik:2013tma}
M.~Dominik, K.~Belczynski, C.~Fryer, D.~E. Holz, E.~Berti, T.~Bulik, I.~Mandel,
  and R.~O'Shaughnessy, ``{Double Compact Objects II: Cosmological Merger
  Rates},'' \href{http://dx.doi.org/10.1088/0004-637X/779/1/72}{{\em Astrophys.
  J.} {\bfseries 779} (2013) 72},
  \href{http://arxiv.org/abs/1308.1546}{{\ttfamily arXiv:1308.1546
  [astro-ph.HE]}}.

\bibitem{Banerjee:2016ths}
S.~Banerjee, ``{Stellar-mass black holes in young massive and open stellar
  clusters and their role in gravitational-wave generation},''
  \href{http://dx.doi.org/10.1093/mnras/stw3392}{{\em Mon. Not. Roy. Astron.
  Soc.} {\bfseries 467} no.~1, (2017) 524--539},
  \href{http://arxiv.org/abs/1611.09357}{{\ttfamily arXiv:1611.09357
  [astro-ph.HE]}}.

\bibitem{Samsing:2013kua}
J.~Samsing, M.~MacLeod, and E.~Ramirez-Ruiz, ``{The Formation of Eccentric
  Compact Binary Inspirals and the Role of Gravitational Wave Emission in
  Binary-Single Stellar Encounters},''
  \href{http://dx.doi.org/10.1088/0004-637X/784/1/71}{{\em Astrophys. J.}
  {\bfseries 784} (2014) 71}, \href{http://arxiv.org/abs/1308.2964}{{\ttfamily
  arXiv:1308.2964 [astro-ph.HE]}}.

\bibitem{Samsing:2017xmd}
J.~Samsing, ``{Eccentric Black Hole Mergers Forming in Globular Clusters},''
  \href{http://dx.doi.org/10.1103/PhysRevD.97.103014}{{\em Phys. Rev. D}
  {\bfseries 97} no.~10, (2018) 103014},
  \href{http://arxiv.org/abs/1711.07452}{{\ttfamily arXiv:1711.07452
  [astro-ph.HE]}}.

\bibitem{Samsing:2018isx}
J.~Samsing and D.~J. D'Orazio, ``{Black Hole Mergers From Globular Clusters
  Observable by LISA I: Eccentric Sources Originating From Relativistic
  $N$-body Dynamics},'' \href{http://dx.doi.org/10.1093/mnras/sty2334}{{\em
  Mon. Not. Roy. Astron. Soc.} {\bfseries 481} no.~4, (2018) 5445--5450},
  \href{http://arxiv.org/abs/1804.06519}{{\ttfamily arXiv:1804.06519
  [astro-ph.HE]}}.

\bibitem{Samsing:2019dtb}
J.~Samsing, D.~J. D'Orazio, K.~Kremer, C.~L. Rodriguez, and A.~Askar,
  ``{Single-single gravitational-wave captures in globular clusters: Eccentric
  deci-Hertz sources observable by DECIGO and Tian-Qin},''
  \href{http://dx.doi.org/10.1103/PhysRevD.101.123010}{{\em Phys. Rev. D}
  {\bfseries 101} no.~12, (2020) 123010},
  \href{http://arxiv.org/abs/1907.11231}{{\ttfamily arXiv:1907.11231
  [astro-ph.HE]}}.

\bibitem{Antonini:2012ad}
F.~Antonini and H.~B. Perets, ``{Secular evolution of compact binaries near
  massive black holes: Gravitational wave sources and other exotica},''
  \href{http://dx.doi.org/10.1088/0004-637X/757/1/27}{{\em Astrophys. J.}
  {\bfseries 757} (2012) 27},
\href{http://arxiv.org/abs/1203.2938}{{\ttfamily arXiv:1203.2938
  [astro-ph.GA]}}.

\bibitem{Hamers:2018hxv}
A.~S. Hamers, B.~Bar-Or, C.~Petrovich, and F.~Antonini, ``{The impact of vector
  resonant relaxation on the evolution of binaries near a massive black hole:
  implications for gravitational wave sources},''
  \href{http://dx.doi.org/10.3847/1538-4357/aadae2}{{\em Astrophys. J.}
  {\bfseries 865} no.~1, (2018) 2},
  \href{http://arxiv.org/abs/1805.10313}{{\ttfamily arXiv:1805.10313
  [astro-ph.HE]}}.

\bibitem{Hoang:2017fvh}
B.-M. Hoang, S.~Naoz, B.~Kocsis, F.~A. Rasio, and F.~Dosopoulou, ``{Black Hole
  Mergers in Galactic Nuclei Induced by the Eccentric Kozai\textendash{}Lidov
  Effect},'' \href{http://dx.doi.org/10.3847/1538-4357/aaafce}{{\em Astrophys.
  J.} {\bfseries 856} no.~2, (2018) 140},
  \href{http://arxiv.org/abs/1706.09896}{{\ttfamily arXiv:1706.09896
  [astro-ph.HE]}}.

\bibitem{Fragione:2019dtr}
G.~Fragione and O.~Bromberg, ``{Eccentric binary black hole mergers in globular
  clusters hosting intermediate-mass black holes},''
  \href{http://dx.doi.org/10.1093/mnras/stz2024}{{\em Mon. Not. Roy. Astron.
  Soc.} {\bfseries 488} no.~3, (2019) 4370--4377},
  \href{http://arxiv.org/abs/1903.09659}{{\ttfamily arXiv:1903.09659
  [astro-ph.GA]}}.

\bibitem{Randall:2017jop}
L.~Randall and Z.-Z. Xianyu, ``{Induced Ellipticity for Inspiraling Binary
  Systems},'' \href{http://dx.doi.org/10.3847/1538-4357/aaa1a2}{{\em Astrophys.
  J.} {\bfseries 853} no.~1, (2018) 93},
\href{http://arxiv.org/abs/1708.08569}{{\ttfamily arXiv:1708.08569 [gr-qc]}}.

\bibitem{Randall:2018nud}
L.~Randall and Z.-Z. Xianyu, ``{An Analytical Portrait of Binary Mergers in
  Hierarchical Triple Systems},''
  \href{http://dx.doi.org/10.3847/1538-4357/aad7fe}{{\em Astrophys. J.}
  {\bfseries 864} no.~2, (2018) 134},
\href{http://arxiv.org/abs/1802.05718}{{\ttfamily arXiv:1802.05718 [gr-qc]}}.

\bibitem{Randall:2018lnh}
L.~Randall and Z.-Z. Xianyu, ``{A Direct Probe of Mass Density Near Inspiraling
  Binary Black Holes},'' \href{http://dx.doi.org/10.3847/1538-4357/ab20c6}{{\em
  Astrophys. J.} {\bfseries 878} no.~2, (2019) 75},
  \href{http://arxiv.org/abs/1805.05335}{{\ttfamily arXiv:1805.05335 [gr-qc]}}.

\bibitem{Silsbee:2016djf}
K.~Silsbee and S.~Tremaine, ``{Lidov-Kozai Cycles with Gravitational Radiation:
  Merging Black Holes in Isolated Triple Systems},''
  \href{http://dx.doi.org/10.3847/1538-4357/aa5729}{{\em Astrophys. J.}
  {\bfseries 836} no.~1, (2017) 39},
\href{http://arxiv.org/abs/1608.07642}{{\ttfamily arXiv:1608.07642
  [astro-ph.HE]}}.

\bibitem{Randall:2019sab}
L.~Randall and Z.-Z. Xianyu, ``{Observing Eccentricity Oscillations of Binary
  Black Holes in LISA},''
\href{http://arxiv.org/abs/1902.08604}{{\ttfamily arXiv:1902.08604
  [astro-ph.HE]}}.

\bibitem{Deme:2020ewx}
B.~Deme, B.-M. Hoang, S.~Naoz, and B.~Kocsis, ``{Detecting
  Kozai\textendash{}Lidov Imprints on the Gravitational Waves of
  Intermediate-mass Black Holes in Galactic Nuclei},''
  \href{http://dx.doi.org/10.3847/1538-4357/abafa3}{{\em Astrophys. J.}
  {\bfseries 901} no.~2, (2020) 125},
  \href{http://arxiv.org/abs/2005.03677}{{\ttfamily arXiv:2005.03677
  [astro-ph.HE]}}.

\bibitem{Nishizawa:2016jji}
A.~Nishizawa, E.~Berti, A.~Klein, and A.~Sesana, ``{eLISA eccentricity
  measurements as tracers of binary black hole formation},''
  \href{http://dx.doi.org/10.1103/PhysRevD.94.064020}{{\em Phys. Rev.}
  {\bfseries D94} no.~6, (2016) 064020},
\href{http://arxiv.org/abs/1605.01341}{{\ttfamily arXiv:1605.01341 [gr-qc]}}.

\bibitem{Nishizawa:2016eza}
A.~Nishizawa, A.~Sesana, E.~Berti, and A.~Klein, ``{Constraining stellar binary
  black hole formation scenarios with eLISA eccentricity measurements},''
  \href{http://dx.doi.org/10.1093/mnras/stw2993}{{\em Mon. Not. Roy. Astron.
  Soc.} {\bfseries 465} no.~4, (2017) 4375--4380},
\href{http://arxiv.org/abs/1606.09295}{{\ttfamily arXiv:1606.09295
  [astro-ph.HE]}}.

\bibitem{Barack:2003fp}
L.~Barack and C.~Cutler, ``{LISA capture sources: Approximate waveforms,
  signal-to-noise ratios, and parameter estimation accuracy},''
  \href{http://dx.doi.org/10.1103/PhysRevD.69.082005}{{\em Phys. Rev.}
  {\bfseries D69} (2004) 082005},
\href{http://arxiv.org/abs/gr-qc/0310125}{{\ttfamily arXiv:gr-qc/0310125
  [gr-qc]}}.

\bibitem{Randall:2019znp}
L.~Randall and Z.-Z. Xianyu, ``{Eccentricity Without Measuring Eccentricity:
  Discriminating Among Stellar Mass Black Hole Binary Formation Channels},''
  \href{http://arxiv.org/abs/1907.02283}{{\ttfamily arXiv:1907.02283
  [astro-ph.HE]}}.

\bibitem{Klein:2015hvg}
A.~Klein {\em et~al.}, ``{Science with the space-based interferometer eLISA:
  Supermassive black hole binaries},''
  \href{http://dx.doi.org/10.1103/PhysRevD.93.024003}{{\em Phys. Rev.}
  {\bfseries D93} no.~2, (2016) 024003},
\href{http://arxiv.org/abs/1511.05581}{{\ttfamily arXiv:1511.05581 [gr-qc]}}.

\bibitem{Yagi:2011wg}
K.~Yagi and N.~Seto, ``{Detector configuration of DECIGO/BBO and identification
  of cosmological neutron-star binaries},''
  \href{http://dx.doi.org/10.1103/PhysRevD.83.044011}{{\em Phys. Rev. D}
  {\bfseries 83} (2011) 044011},
  \href{http://arxiv.org/abs/1101.3940}{{\ttfamily arXiv:1101.3940
  [astro-ph.CO]}}. [Erratum: Phys.Rev.D 95, 109901 (2017)].

\bibitem{Wen:2002km}
L.~Wen, ``{On the eccentricity distribution of coalescing black hole binaries
  driven by the Kozai mechanism in globular clusters},''
  \href{http://dx.doi.org/10.1086/378794}{{\em Astrophys. J.} {\bfseries 598}
  (2003) 419--430}, \href{http://arxiv.org/abs/astro-ph/0211492}{{\ttfamily
  arXiv:astro-ph/0211492}}.

\bibitem{Randall:future}
N.~DePorzio, L.~Randall, A.~Shelest, and Z.-Z. Xianyu , to appear.

\bibitem{Martinez:2020lzt}
M.~A.~S. Martinez {\em et~al.}, ``{Black Hole Mergers from Hierarchical Triples
  in Dense Star Clusters},''
  \href{http://dx.doi.org/10.3847/1538-4357/abba25}{{\em Astrophys. J.}
  {\bfseries 903} no.~1, (2020) 67},
  \href{http://arxiv.org/abs/2009.08468}{{\ttfamily arXiv:2009.08468
  [astro-ph.GA]}}.

\bibitem{Peters:1964zz}
P.~C. Peters, ``{Gravitational Radiation and the Motion of Two Point Masses},''
\href{http://dx.doi.org/10.1103/PhysRev.136.B1224}{{\em Phys. Rev.} {\bfseries
  136} (1964) B1224--B1232}.

\end{thebibliography}
\end{document}